\documentclass[final,5p,times]{elsarticle}

\usepackage{amssymb}
\usepackage{amsthm}

\usepackage{lmodern}
\usepackage{amssymb,amsmath,xfrac}
\usepackage{graphicx,subcaption,wrapfig,float}
\usepackage{booktabs} 	
\usepackage{longtable} 	
\usepackage{multirow} 	
\usepackage{siunitx} 	
\usepackage{pifont,amssymb}
\usepackage{csquotes}

\usepackage{mathabx}
\usepackage{mathtools}
\usepackage{enumitem}
\usepackage{wasysym}

\usepackage{xparse}%
\usepackage{blindtext}
\usepackage{morewrites}
\usepackage{xkeyval}%

\usepackage{caption}
\usepackage{wrapfig}

\setlist[itemize]{leftmargin=1mm,itemindent=4mm,noitemsep}
\setlist[enumerate,1]{label=\alph*),font=\itshape,leftmargin=1mm,itemindent=4mm,noitemsep,nolistsep}

\newcommand{\cmark}{\ding{51}}%
\newcommand{\xmark}{\ding{55}}%
\newcommand{\comment}[1]{}

\makeatletter
\def\old@comma{,}
\catcode`\,=13
\def,{%
  \ifmmode%
    \old@comma\discretionary{}{}{}%
  \else%
    \old@comma%
  \fi%
}
\makeatother

\allowdisplaybreaks

\begin{document}
\begin{frontmatter}



\title{HUAP: Practical Attribute-based Access Control Supporting Hidden Updatable Access Policies for Resource-Constrained Devices\tnoteref{now}}
\tnotetext[now]{This paper is an extension of work presented in ISCISC 2019 \cite{39}.}


\author[SharifEE]{Mostafa Chegenizadeh\corref{cor1}}
\cortext[cor1]{Corresponding author.}
\ead{mostafa.chegenizadeh@alum.sharif.edu}
\author[Amirkabir]{Mohammad Ali}
\ead{mali71@aut.ac.ir}
\author[SharifRI]{Javad Mohajeri}
\ead{mohajer@sharif.ir}
\author[SharifEE]{Mohammad Reza Aref}
\ead{aref@sharif.ir}

\address[SharifEE]{Dept. of Electrical Engineering Sharif University of Technology Tehran, Iran}
\address[Amirkabir]{Dept. of Mathematics and Computer Science Amirkabir University of Technology Tehran, Iran}
\address[SharifRI]{Electronic Research Institute Sharif University of Technology Tehran, Iran}


\begin{abstract}
Attribute-based encryption (ABE) is a promising cryptographic mechanism for providing confidentiality and fine-grained access control in the cloud-based area.
However, due to high computational overhead, common ABE schemes are not suitable for resource-constrained devices.
Moreover, data owners should be able to update their defined access policies efficiently, and in some cases, applying hidden access policies is required to preserve the privacy of clients and data.
In this paper, we propose a ciphertext-policy attribute-based access control scheme which for the first time provides online/offline encryption, hidden access policy, and access policy update simultaneously.
In our scheme, resource-constrained devices are equipped with online/offline encryption reducing the encryption overhead significantly.
Furthermore, attributes of access policies are hidden such that the attribute sets satisfying an access policy cannot be guessed by other parties.
Moreover, data owners can update their defined access policies while outsourcing a major part of the updating process to the cloud service provider.
In particular, we introduce blind access policies that enable the cloud service provider to update the data owners' access policies without receiving a new re-encryption key.
Besides, our scheme supports fast decryption such that the decryption algorithm consists of a constant number of bilinear pairing operations.
The proposed scheme is proven to be secure in the random oracle model and under the hardness of Decisional Bilinear Diffie–Hellman (DBDH) and Decision Linear (D-Linear) assumptions.
Also, performance analysis results demonstrate that the proposed scheme is efficient and practical.
\end{abstract}

\comment{
\begin{graphicalabstract}
\end{graphicalabstract}
}

\comment{
\begin{highlights}
\item HUAP achieves online/offline encryption, hidden access policy, access policy update.
\item HUAP utilizes blind access policy for policy updating without new re-encryption key.
\item HUAP is proven to be secure under the DBDH assumption and the D-Linear assumption.
\end{highlights}
}


\end{frontmatter}


\section{INTRODUCTION}
With rapidly increasing the number of cloud-based services, the need for methods to provide data secrecy and user privacy grows significantly \cite{1}. Cloud computing technology enables data owners to outsource their private data to a cloud service provider and define an access policy preventing unauthorized parties from accessing their data.

Attribute-based encryption (ABE) \cite{7} offers access control for protecting information within the cloud computing environment. ABEs are divided into two primary categories key-policy ABE (KP-ABE) \cite{8} and ciphertext-policy ABE (CP-ABE) \cite{9}.
In a KP-ABE scheme, access rights of users are determined by a trusted third party, and ciphertexts are labeled by some attributes. A user can decrypt a ciphertext if and only if the attributes of the ciphertext satisfy the user's access right.
However, in a CP-ABE scheme, access rights of users are specified according to their attributes, and each ciphertext is associated with an access policy such that only users whose attributes satisfy the access policy can recover the associated message \cite{41}. As in CP-ABE data owners can determine the privileges of authorized users, it is more suitable for real cloud-based applications like smart health (s-health) \cite{14}.

Although CP-ABE brings great benefits, there are also some main challenges. Firstly, in traditional CP-ABE, access policies are stored in a clear-text form. As access policies consist of authorized users' attributes, revealing the access policies may leak some sensitive information about the associated data or the associated recipients. Anonymous ABE (A-ABE) schemes \cite{14,5,10,11,12,13,15,16} alleviate this problem by affording hidden access policies. Indeed, in these schemes, no party can obtain any information about the authorized users' attributes.

Secondly, in many situations, data owners need to update their defined access policies, revoke the access right of some data users (policy deletion), or grant some new access privileges to some other users (policy addition). The revoked users must be unable to extract the underlying values that are encrypted under the new access policies. An obvious solution to this problem is to decrypt and then re-encrypt the data. However, it is clearly impractical for large amounts of data. To efficiently address the problem, data owners should be able to outsource the updating process to a proxy server. However, in traditional ciphertext-policy attribute-based proxy re-encryption (CP-ABPRE) schemes \cite{18,19,20,24,31}, data owners have to generate some re-encryption keys whenever they need a policy update. As a result, growing the number of ciphertexts as well as rising the number of access policy updates, makes the updating process inefficient \cite{2}. Moreover, as we know, the existing CP-ABPRE schemes require that the data owner be online to generate the re-encryption key, while the data owner may not be available when the access policy update is needed, for example, due to limited network bandwidth or limited computational power \cite{3}. Therefore, the process of access policy update should be feasible even when the data owner is offline.

Thirdly, in many existing applications like s-health, data owners usually use resource-limited devices for encrypting and sending data to the cloud service provider. Therefore, the data owners have trouble in completing the whole computations of the encryption algorithm \cite{40}. Online/offline encryption mechanism \cite{27,28,29,30} is a promising solution to this problem. In this setting, the encryption process is divided into two phases: offline phase and online phase.
In the offline phase, the device can access enough power resources and has enough time to generate some offline ciphertexts while messages are not known. In the online phase, while the device can access limited power and computational resources, once a message is known the device uses a pre-computed offline ciphertext to obtain an online ciphertext in a short period of time \cite{43}.

Fourthly, in traditional CP-ABE, the same entity collects data and also defines access policies. However, in reality, there may be several devices that collect data while another party defines the access policy.
Directly adopting traditional CP-ABE in such a situation requires that all of the data collector devices be aware of the current defined access policy and encrypt data according to it. Therefore, whenever the corresponding data owner wants to define a new access policy over the data, the encryption algorithm running by these devices needs to be updated, while re-programming these devices is difficult in some applications like s-health, and hence changing the encryption algorithm is not feasible \cite{42}.
Moreover, by adopting traditional CP-ABE, all of the data collector devices should share a similar set of secret parameters. Therefore, revealing secret parameters of each of these devices threatens the security of all the others.

\begin{figure}[t]
    \centering
        \framebox[1.01\linewidth]{
            \parbox{1\linewidth}{
            \begin{subfigure}{1\linewidth}
                \includegraphics[width=1\linewidth]{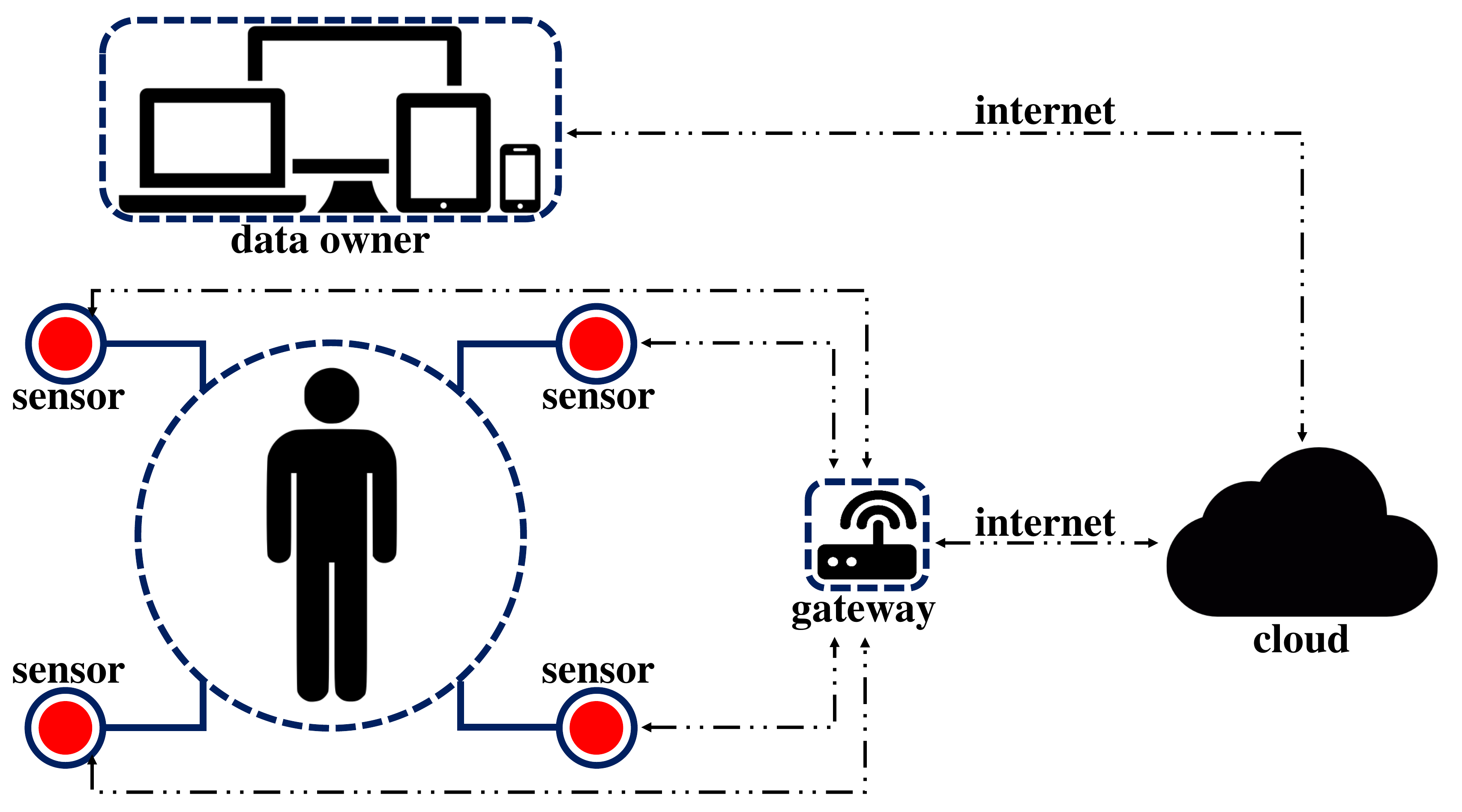}
            \end{subfigure}
        }
    }
    \caption{Body Sensor Network.}
    \label{fig:BSN}
\end{figure}

To make sense, consider the following s-health scenario in which simultaneously resolving all of the above issues is necessary. Main entities in a Body Sensor Network (BSN) are shown in Fig. \ref{fig:BSN}. In a BSN, there are several resource-constrained sensors that collect health data from a patient's body, where each sensor collects a specific kind of data such as blood pressure, blood oxygen level, heart rate, respiratory rate, body temperature, etc. Each sensor encrypts its collected data independently and then outsources the encrypted data to the cloud through a gateway.
At the other side, the data owner connects to the cloud and defines an access policy for the whole of the outsourced health data. In this case, the following security and performance requirements should be fulfilled:
\begin{enumerate}[label=\arabic*)]
    	\item To preserve the attribute privacy of authorized data users, the attribute sets satisfying the access policy should be hidden.
    	\item The data owner should be able to update the defined access policy efficiently.
    	\item The computational overhead on the resource-constrained sensors should be as low as possible.
    	\item The sensors should encrypt the collected data independent of the defined access policy. Therefore, updating the access policy should not change the performance of these sensors. Moreover, The sensors should work independently such that revealing secret parameters of a sensor does not threaten the security of the data collected by the other sensors.\\
\end{enumerate}

	In this paper, to address the aforementioned challenges we present the HUAP scheme that provides a secure fine-grained access control system for resource-constrained devices in cloud-based applications.
The contributions of this work can be summarized as follows:
\begin{itemize}
	\item[$\bullet$] In HUAP, access policies are hidden, and hence the attribute privacy of the authorized data users is preserved such that the attribute sets satisfying the defined access policy cannot be guessed by unauthorized data users or the cloud service provider.
        \item[$\bullet$] HUAP realizes online/offline encryption in order to reduce the encryption overhead.
        Moreover, the proposed scheme achieves fast decryption, where the decryption algorithm consists of a constant number of bilinear pairing operations.
        Therefore, the computational overhead of encryption and decryption is considerably decreased.
        \item[$\bullet$] HUAP achieves a large attribute universe, where any string can be used as an attribute while the number of public parameters of the system remains constant.
        \item[$\bullet$] HUAP introduces a new concept called blind access policy. The attribute sets that satisfy a blind access policy are determined by the associated data owner defining the blind access policy. On the other side, the cloud service provider can encrypt several messages under a pre-generated blind access policy without knowing anything about the associated attributes that satisfy the policy.
        \item[$\bullet$] HUAP enables data owners to efficiently update their defined access policies. To reduce computational overhead at the user side, most of the operations related to the access policy update process are outsourced to the cloud service provider, without leakage of any information about the previous and new access policies.
        In particular, a data owner can remain offline during the process of policy deletion, as the cloud can update the access policy without receiving any new re-encryption key.
        \item[$\bullet$] In HUAP, data collector devices perform independent of the access policy defined by the corresponding data owner.
        As a result, updating the access policy does not affect the performance of these devices.
		Moreover, many devices can perform simultaneously to collect the data corresponding to a data owner such that revealing information about the secret parameters of a device does not threaten the security of the other devices, privacy of their collected data, or hiddenness of the defined access policy.
        \item[$\bullet$] We prove that HUAP is selective ciphertext-policy and chosen-plaintext secure (CPA-secure) under the Decisional Bilinear Diffie–Hellman (DBDH) assumption and the Decisional Linear (DL) assumption in random oracle model.
\end{itemize}

\section{RELATED WORK}
In this section, we summarize the related work on attribute-based encryption, anonymous attribute-based encryption, updating access policy in attribute-based encryption, and online/offline cryptography.

\emph{\textbf{Attribute-based encryption.}} After introducing the notion of attribute-based encryption (ABE) by Sahai and Waters \cite{7}, key-policy attribute-based encryption (KP-ABE) proposed by Goyal et al. \cite{8}, and Ciphertext-policy attribute-based encryption (CP-ABE) proposed by Bethencourt et al. \cite{9}, divided this class of cryptographic schemes into two primary groups. However, CP-ABE seems to be more suitable than KP-ABE for providing fine-grained access control in public cloud-based data sharing applications. Because, in CP-ABE, data owners can enforce their desired access policies over their outsourced data, while in KP-ABE, this is the attribute authority that encapsulates the access policies in secret keys issued for data users, and the data owners can only define a set of attributes related to their outsourced data.
We refer the reader to \cite{37,38} to study more about the topic.

\emph{\textbf{Anonymous attribute-based encryption.}} Anonymous attribute-based encryption (A-ABE) has been proposed to protect the users' attribute privacy. In anonymous ABE schemes, to protect sensitive information included in access policies of ciphertexts, the policies are hidden such that an unauthorized data user whose attributes do not satisfy an access policy cannot guess which attributes are required to decrypt the associated ciphertext.
With regard to hidden access policies, there are two main categories in the literature: fully hidden and partially hidden.
In fact, access policies consist of a set of attributes expressed as a couple: attribute name and attribute value \cite{45}.
To be specific, a fully hidden access policy obscures the names of the attributes in the policy as well as the values associated with each attribute name. On the other hand, in a partially hidden access policy only the attribute values are hidden and the secrecy of the attribute names is not provided.
Kapadia et al. \cite{10} proposed the first anonymous ABE scheme which supports AND-gate access policies on positive and negative attributes, but their scheme was vulnerable to collusion attacks.
Nishide et al. \cite{11} designed an efficient anonymous ABE scheme resisting collusion attacks.
Li et al. \cite{12} proposed an anonymous ABE scheme to realize user accountability.
Afterward, Lai et al. \cite{13} proposed an anonymous ABE scheme to protect user privacy and achieve full security. However, in their scheme, data users have to repeat the decryption algorithm until successful decryption is achieved, and if all the possible decryption tests are unsuccessful, then the user concludes that his attributes do not satisfy the underlying access policy. It is obvious that this approach is time-consuming and the scheme is inefficient.
To address the problem, Zhang et al. \cite{5} designed a technique called match-then-decrypt that enables the data users to efficiently check whether their attributes satisfy a hidden access policy or not. Subsequently, they proposed another anonymous ABE scheme that also supports large universe and linear secret sharing scheme (LSSS) policies \cite{14}.
However, their proposed scheme is not adequately efficient as it is based on composite order groups dealing with large elements.
Hao et al. \cite{15} realized a fuzzy attribute positioning mechanism that fully hides access policies by applying garbled bloom filter.
Xiong et al. \cite{32} proposed an anonymous attribute-based broadcast encryption scheme in edge computing that realizes direct revocation by embedding the list of identities of authorized data users in the ciphertext.
However, by raising the number of users in the system, the number of system public parameters grows, and hence the scheme is not suitable for large networks.
Zhang et al. \cite{16} proposed an anonymous ABE scheme for personal health record systems. Their proposed scheme supports fast decryption. Also, by using hash functions, it enables data users to verify the validity of the received ciphertext.
However, none of the aforementioned schemes support access policy update.

\emph{\textbf{Online/offline cryptography.}} The notion of online/offline was first formalized by Even et al. \cite{26} in digital signatures.
In an online/offline signature scheme, the offline phase is performed before the message is known. Once the message is determined, the data owner uses a trapdoor to generate a dual signature.
The technique of online/offline ABE was introduced by Hohenberger et al. \cite{27}.
Datta et al. \cite{28} proposed the first adaptive payload-hiding online/offline KP-ABE scheme which supports a large attribute universe.
Liu et al. \cite{29} proposed an online/offline CP-ABE scheme for resource-constrained devices in the mobile cloud computing area.
Li et al. \cite{30} proposed an online/offline KP-ABE scheme that moves a vast majority of the encryption computational overhead on the data owner’s side to the offline phase.
The scheme realizes public ciphertext test before performing the decryption algorithm, and also eliminates a major part of the computational operations by adding some public parameters to the system.
However, the aforementioned schemes support neither access policy update nor hidden access policies.

\emph{\textbf{Access policy update.}} Updating the access policy is one of the most critical and essential tasks for access control administration.
According to the existing schemes, the approach for access policy updating can be divided into the following:
1) deploying proxy re-encryption, 2) embedding required update parameters in the ciphertext.
The notion of proxy re-encryption (PRE) was first formalized by Blaze et al. \cite{17}.
The first ciphertext-policy attribute-based proxy re-encryption (CP-ABPRE) scheme was proposed by Liang et al. \cite{18}.
In their cloud-based access control system, deploying CP-ABE, data owners can generate a re-encryption key to outsource updating their defined access policies.
Using this re-encryption key, the proxy server updates the access policy of a ciphertext.
Subsequently, Luo et al. \cite{19} proposed another CP-ABPRE scheme supporting multi-value positive attributes.
Afterward, an efficient CP-ABPRE scheme with a constant number of pairing operations was proposed by Seo et al. \cite{20}.
Liu et al. \cite{6,21} proposed the notion of time-based proxy re-encryption in which the access policies and the attribute secret keys are updated with respect to the global time of the system.
Li et al. \cite{22} proposed a fine-grained access control scheme with policy updating for the smart grid area.
Also, Jiang et al. \cite{3} designed a CP-ABE scheme supporting access policy update based on AND-gate access policies.
Huang et al. \cite{23} proposed a hierarchical ABE for resource-constrained IoT devices that supports updating the access policies.
In order to relieve the local computational burden, their scheme partially outsources the process of computationally expensive encryption operations to a gateway, and decryption operations to the cloud.
Li et al. \cite{24} proposed a CP-ABE scheme that enables the data owner to outsource updating the access policy and also the shared files to reduce the storage and communication costs of the client.
Sethi et al. \cite{25} constructed a multi-authority ABE scheme that supports white-box traceability and access policy update.
Recently, Belguith et al. \cite{2} have presented a KP-ABE scheme that verifiably outsources data decryption process to edge nodes.
In their scheme, the data owner sends some secret parameters along with the ciphertext to the cloud. The cloud can utilize these parameters to update the access policy of the ciphertext.
Hence, the scheme is capable of offline policy deletion, where the cloud can update the access policy without receiving any new re-encryption key from the data owner. However, none of the aforementioned schemes hides the access policy.

\begin{table*}[t]
  \centering 
  \vspace{-5mm}
    \caption{SECURITY AND PERFORMANCE COMPARISON.}
    \label{tab:table_functionality_compare}
    \setlength
    \extrarowheight{-5pt}
    \resizebox{0.9\linewidth}{!}{%
        \begin{tabular}{l l l l l l l l}
          \toprule 
          \multicolumn{1}{c}{\textbf{Scheme}} &
          \multicolumn{1}{l}{
            \begin{tabular}{c}
              \textbf{Access Policy}\\
              \textbf{Type}\\
            \end{tabular}
          } &
          \multicolumn{1}{l}{
            \begin{tabular}{c}
              \textbf{Hidden}\\
              \textbf{Access Policy}\\
            \end{tabular}
          } &
          \multicolumn{1}{l}{
            \begin{tabular}{c}
              \textbf{Large}\\
              \textbf{Universe}\\
            \end{tabular}
          } &
          \multicolumn{1}{l}{
            \begin{tabular}{c}
              \textbf{Fast}\\
              \textbf{Decryption}\\
            \end{tabular}
          } &
          \multicolumn{1}{l}{
            \begin{tabular}{c}
              \textbf{Online/offline}\\
              \textbf{Encryption}\\
            \end{tabular}
          } &
          \multicolumn{1}{l}{
            \begin{tabular}{c}
              \textbf{Access}\\
              \textbf{Policy Update}\\
            \end{tabular}
          } &
          \multicolumn{1}{l}{
            \begin{tabular}{c}
              \textbf{Offline Policy}\\
              \textbf{Deletion}\\
            \end{tabular}
          }\\
          \midrule 
                \cite{14} &   LSSS	   & \cmark\ Partially     &\cmark &\xmark &\xmark &\xmark &\textbf{-}	\\
                \cite{5}  & AND-gates  & \cmark\ Fully	       &\cmark &\cmark &\xmark &\xmark &\textbf{-}	\\
                \cite{15} &   LSSS	   & \cmark\ Fully	       &\xmark &\xmark &\xmark &\xmark &\textbf{-}	\\
                \cite{16} &   LSSS	   & \cmark\ Partially     &\cmark &\cmark &\xmark &\xmark &\textbf{-}	\\
                \cite{24} &   LSSS	   & \xmark\	           &\xmark &\xmark &\xmark &\cmark &\xmark		\\
                \cite{31} & AND-gates  & \cmark\ Fully	       &\xmark &\xmark &\xmark &\cmark &\xmark		\\
                \cite{2}  &   LSSS	   & \xmark\	           &\xmark &\xmark &\xmark &\cmark &\cmark		\\
                \cite{3}  & AND-gates  & \xmark\	           &\xmark &\cmark &\xmark &\cmark &\cmark		\\
                \cite{29} &   LSSS	   & \xmark\	           &\xmark &\xmark &\cmark &\xmark &\textbf{-}	\\
                \cite{30} &   LSSS	   & \xmark\	           &\xmark &\xmark &\cmark &\xmark &\textbf{-}	\\
                \cite{45} &   LSSS	   & \cmark\ Partially     &\xmark &\xmark &\xmark &\cmark &\xmark		\\
                \cite{32} &   LSSS	   & \cmark\ Partially     &\cmark &\xmark &\xmark &\xmark &\textbf{-}		\\
                \cite{23} &   LSSS	   & \xmark\	           &\xmark &\xmark &\xmark &\cmark &\xmark		\\
                \cite{25} &   LSSS	   & \xmark\	           &\cmark &\xmark &\xmark &\cmark &\xmark		\\
                \cite{46} &   LSSS	   & \cmark\ Partially	   &\cmark &\xmark &\cmark &\xmark &\textbf{-}		\\
                \cite{47} &   LSSS	   & \cmark\ Partially	   &\cmark &\xmark &\cmark &\xmark &\textbf{-}		\\
                \cite{48} &   AND-gates	   & \cmark\ Fully	   &\xmark &\cmark &\cmark &\xmark &\textbf{-}		\\
                HUAP      & AND-gates  & \cmark\ Fully	       &\cmark &\cmark &\cmark &\cmark &\cmark		\\
          \bottomrule 
        \end{tabular}
    }
\end{table*}

To simultaneously support anonymity and access policy update, Zhang et al. \cite{31} proposed an anonymous CP-ABPRE scheme in which the proxy server can update hidden access policies.
However, to update an access policy, an authorized data user should generate a new re-encryption key for the proxy server, while the data owner cannot generate a valid re-encryption key.
In addition, generating re-encryption keys requires running the whole of the encryption algorithm which increases the computational and communication overhead on the user side. Moreover, it is necessary for the authorized data user to be online while providing the required re-encryption key.
Afterward, Yan et al. \cite{45} proposed a multi-authority attribute-based encryption scheme with dynamic policy updating for personal health record systems. Their scheme uses partially hidden access policies to protect the user's identity and attribute privacy. However, it does not fully hide the attributes in access policies and hence the attribute names are disclosed.

On the other hand, some other schemes have been proposed to simultaneously support anonymity and online/offline encryption.
Yan et al. \cite{46} proposed an attribute-based encryption scheme with partially hidden policies for the Internet of Things. In this scheme, data users can outsource the decryption process to the cloud and then verify returned results. However, their construction is based on inefficient composite-order groups.
Tian et al. \cite{47} proposed a multi-authority attribute-based access control scheme with partially hidden policies for intelligent transportation systems. This scheme supports online/offline encryption and outsourced decryption to achieve lightweight computation for IoT devices.
Sun et al. \cite{48} proposed a lightweight policy-hiding attribute-based access control scheme with online/offline encryption for IoT-oriented s-health applications. The authors in this scheme propose an optimized vector transformation approach to decrease the overhead of key generation, encryption, and decryption algorithms. However, access policies are AND-gates on positive and negative attributes with wildcards, and hence the scheme is less expressive than other relevant schemes.

Table \ref{tab:table_functionality_compare} summarizes the result of functional comparison between our proposed scheme and other similar ABE schemes in the literature that support at least one of the following features: 1) hidden access policy, 2) online/offline encryption, 3) access policy update.

This paper is an extended version of a conference paper published in \cite{39}. We extend our previous work by expanding system architecture, improving the related cryptographic structures, evaluating performance based on the actual execution time, and providing security proof in detail.

\section{PRELIMINARIES}

In this section, we briefly present  some cryptographic notions related to our work.

\subsection{Cryptographic Background}

\begin{enumerate}
    \item \textit{Bilinear pairing:} Assume that $\mathbb{G}$ and $\mathbb{G}_T$ are two cyclic multiplicative groups of a large prime order $p$, $1_\mathbb{G}$ is the identity of $\mathbb{G}$, $1_{\mathbb{G}_T}$ is the identity of $\mathbb{G}_T$, and $g$ is a generator of $\mathbb{G}$. The map $e:\mathbb{G}\times \mathbb{G}\rightarrow \mathbb{G}_T$ is a bilinear pairing, if it satisfies the following properties:
\begin{enumerate}
            \item \textit{Bilinear:} For any $a,b\in Z_p$, we have $e\left(g^a,g^b\right)={e(g,g)}^{a.b}$.
            \item \textit{Non-degenerate:} There exists at least two $g_1,g_2\in\mathbb{G}$ such that $e(g_1,g_2)\neq1_{\mathbb{G}_T}$.
            \item \textit{Computable:} For all $g_1,g_2\in\mathbb{G}$ ,$e(g_1,g_2)$ can be computed by a polynomial-time algorithm.
	\end{enumerate}

    \item \textit{Proxy re-encryption:} Usually a proxy re-encryption scheme consists of three polynomial time algorithms: key generation, encryption and re-encryption, and three main entities: $Alice$, $Bob$, and a $proxy$. At first, there is a message $M$ encrypted by $Alice$’s public key ${PK}_{Alice}$ noted as $\ C_{Alice}$. Then a re-encryption key ${RK}_{Alice\rightarrow Bob}$ is sent to the $proxy$ by $Alice$. The re-encryption key enables the $proxy$ to re-encrypt $C_{Alice}$ and create a new ciphertext $C_{Bob}$ that is encrypted by ${PK}_{Bob}$. The main challenge in the proxy re-encryption is preventing the $proxy$ from obtaining any information about the message $M$, and secret keys of $Alice$ and $Bob$.

\end{enumerate}

\subsection{Complexity Assumptions}
\label{section:complexity_assumptions}

\begin{itemize}
    \item[$$] \emph{1) Decisional Bilinear Diffie–Hellman (DBDH) assumption:} Let $\mathbb{G}$ be a cyclic multiplicative group of a large prime order $p$, $g$ be a generator of $\mathbb{G}$, $e:\mathbb{G}\times\mathbb{G}\rightarrow\mathbb{G}_T$ be a bilinear pairing, $Z\in_R\mathbb{G}_T$, and $a,b,c\in_RZ_p$. We say that the DBDH assumption \cite{33} holds if no probabilistic polynomial-time algorithm can distinguish the tuple $[g,g^a,g^b,g^c,{e(g,g)}^{abc}]$ from the tuple $[g,g^a,g^b,g^c,Z]$ with non-negligible advantage.

    \item[$$] \emph{2) Decision Linear (D-Linear) assumption:} Let $\mathbb{G}$ be a cyclic multiplicative group of a large prime order $p$, $g$ be a generator of $\mathbb{G}$, and $z_1,z_2,z_3,z_4,z\in_RZ_p$. We say that the D-Linear assumption \cite{34} holds if no probabilistic polynomial-time algorithm can distinguish the tuple $[g,\ g^{z_1},g^{z_2},g^{z_1z_3},g^{z_2z_4},g^{z_3+z_4}]$ from the tuple $[g,g^{z_1},g^{z_2},g^{z_1z_3},g^{z_2z_4},g^z]$ with non-negligible advantage.

\end{itemize}

\subsection{Aceess Policies}

Access policy is a rule $W$ over some attributes. For a given attribute list $L$, access policy returns true if $L$ satisfies $W$ and the notation $L\models W$ represents this situation. Otherwise, if $L$ does not satisfy $W$, the notation $L\nvDash W$ is used and access policy returns false.

In our scheme, the access policies consist of multiple AND-gates supporting multi-value attributes and wildcards where wildcard $\ast$ is known as \enquote{don't care} value.
The notion generalizes the common concept of access policies in\cite{11} consisting of a single AND-gate supporting multi-value attributes and wildcards. Assume that $n$ is the total number of attributes in the system and $\mathbb{U}=\{\omega_1,\omega_2,\ldots,\omega_n\}$ is the universal attribute set. Each attribute can take multiple values and the set of possible values for $\omega_i$ is $S_i=\{v_{i,1},v_{i,2},\ldots,v_{i,n_i}\}$ where $n_i$ is the number of possible values for $\omega_i$, $i=1,2,\ldots,n$.

Given an attribute list $L=\left[L_1,L_2,\ldots,L_n\right]$ and an access policy $A=\bigvee_{j=1}^{m}W_j$, where $ W_j=\left[W_{j,1},W_{j,2},\ldots,W_{j,n}\right]$ and for all $1\le i\le n, W_{j,i}\subseteq S_i$ and $L_i\in S_i$. In particular, $W_{j,i}=\ast$ means that $W_{j,i}=S_i$.
We say that $L$ satisfies $W_j$ and we write $L\models W_j$, if $L_i\in W_{j,i}$ for all $1\le i\le n$. Otherwise, we say it does not satisfy $W_j$, $L\nvDash W_j$.
Also, a given attribute list $L=\left[L_1,L_2,\ldots,L_n\right]$ satisfies an access policy $A=\bigvee_{j=1}^{m}W_j$ if $L\models W_j$ for some $1\le j\le m$. Otherwise, $L$ does not satisfy $A$.

For example, assume that there are five attributes in the universe. We consider an access policy $A=W_1\bigvee W_2$, where $W_1=[W_{1,1}=\{v_{1,1},v_{1,3}\},W_{1,2}=\{v_{2,1},v_{2,2}\},W_{1,3}=\ast,W_{1,4}=\ast,W_{1,5}=\ast]$, and $W_2=[W_{2,1}=\ast,W_{2,2} =\{v_{2,4}\},W_{2,3}=\{v_{3,2}\},W_{2,4}=\ast,W_{2,5}=\ast]$. According to the above access policy, if a recipient wants to decrypt a message corresponding to  $A$, he must have the value $v_{1,1}$ or $v_{1,3}$ for $\omega_1$, and $v_{2,1}$ or $v_{2,2}$ for $\omega_2$, while the values for $\omega_3$, $\omega_4$, and $\omega_5$ are not cared for, or he has to have the value $v_{2,4}$ for $\omega_2$, and $v_{3,2}$ for $\omega_3$, while the values for $\omega_1$, $\omega_4$, and $\omega_5$ are not cared for.

\begin{figure}[t]
    \centering
            \parbox{1\linewidth}{
            \begin{subfigure}{1\linewidth}
                \includegraphics[width=1\linewidth]{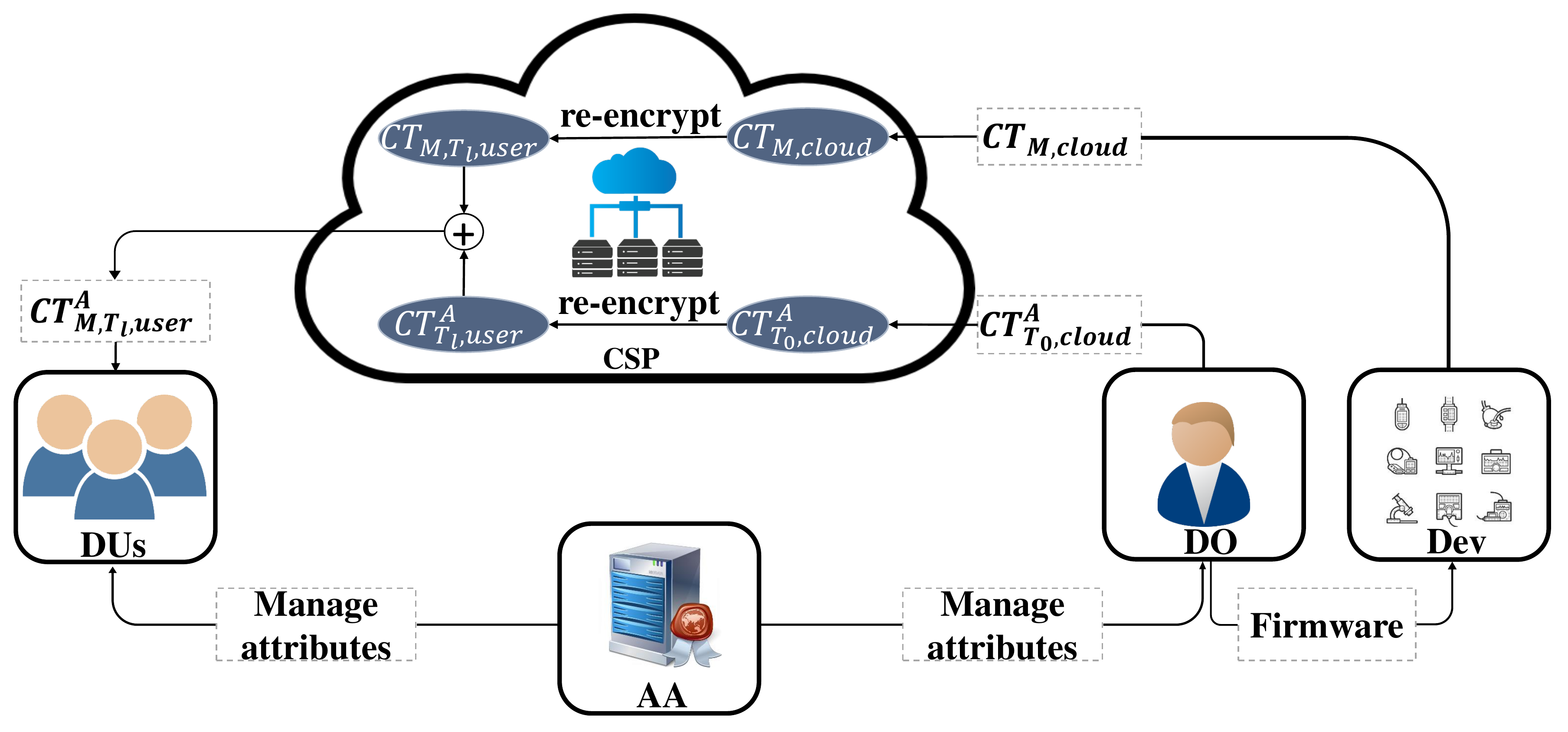}
            \end{subfigure}
        }

    \caption{HUAP Architecture.}
    \label{fig:Architecture}
\end{figure}

\section{SYSTEM MODEL AND DESIGN GOALS}
\label{section:SYSTEM_MODEL}

In this section, firstly we present the architecture of the system. Then, we give an overview of the proposed scheme. The later consists of two parts: the algorithms of the scheme and the flow of sharing data in the system. Afterward, we review the trust model and security assumptions, and finally we describe the design goals of the proposed scheme.

\subsection{System Architecture}
As shown in Fig. \ref{fig:Architecture}, the system architecture of the proposed scheme consists of the following entities:

\begin{itemize}
        \item[$\bullet$] \textbf{Attribute Authority (AA):} It is a fully trusted entity which generates system public key and system master key. It also generates attribute secret keys of users.
        \item[$\bullet$] \textbf{Cloud Service Provider (CSP):} It is an honest but curious entity with abundant storage capacity and computational power. Data encrypted by DOs are stored and managed by the CSP. It also provides the fine-grained access control service.
        \item[$\bullet$] \textbf{Device (Dev):} It is a device that generates private messages, encrypts, and sends them to the CSP.
        \item[$\bullet$] \textbf{Data Owner (DO):} It is a user that wishes to define a hidden access policy over encrypted messages generated by Dev and outsourced to the CSP.
        \item[$\bullet$] \textbf{Data User (DU):} It is a user with an attribute secret key associated with an attribute list $L$. It aims to access some encrypted data outsourced into the CSP. DU can decrypt the encrypted data if and only if her/his attributes satisfy the access policy of the ciphertext.
\end{itemize}

In this architecture, we have assumed that Dev is connected directly to the CSP. However, in practice, this connection can be through a semi-trusted gateway which honestly relays ciphertexts to the CSP\cite{44}.

\subsection{Overview of Scheme}

The proposed HUAP scheme consists of the following algorithms. The most relevant notations used in our scheme are summarized in Table \ref{tab:notations}:

\begin{table*}[t]
  \centering 
  \vspace{-5mm}  
    \caption{SUMMARY OF NOTATIONS.}
    \label{tab:notations}
    \setlength
    \extrarowheight{-5pt}
    \resizebox{0.75\textwidth}{!}{
        \begin{tabular}{l l}
          \toprule 
          \textbf{Notation} & \textbf{Description}\\
          \midrule 
			$PK$ & system public key\\
			$MK$ & system master key\\
			${SK}_L$ & attribute secret key (associated with attribute list $L$)\\
			$PP$ & data public parameter\\
			$SP$ & data secret parameter\\
			$RK$ & re-encryption key\\
			${CT}_{off}$ & offline ciphertext\\
			${CT}_{M,cloud}$ & message ciphertext\\
			${CT}_{T_0,cloud}^A$ & policy ciphertext (associated with hidden access policy $A$)\\
			${CT}_{cloud}^A$ & cloud ciphertext (associated with hidden access policy $A$)\\
			${CT}_{M,{T_l},user}$ & re-encrypted message ciphertext (with respect to timestamp $T_l$)\\
			${CT}_{{T_l},user}^A$ & re-encrypted policy ciphertext (associated with hidden access policy $A$ with respect to timestamp $T_l$)\\
			${CT}_{M,T_l,user}^A$ & user ciphertext (associated with hidden access policy $A$ with respect to timestamp $T_l$)\\
			${dk}_{T_l}$ & data decryption key (with respect to timestamp $T_l$)\\
			${CT}_{W_j,T_l}$ & data decryption key $dk_{T_l}$ encrypted under sub-policy $W_j$ (with respect to timestamp $T_l$)\\
			${\widetilde{CT}}_{W_j,T_0}$ & blind access policy associated with sub-policy $W_j$\\
          \bottomrule 
        \end{tabular}
    }
\end{table*}

\begin{enumerate}[label=\arabic*)]
    \item $SystemSetup(1^\lambda)\rightarrow(PK,MK)$: The system setup algorithm is run by AA. A security parameter $\lambda$ is chosen as the input of the algorithm. The outputs of the algorithm are the system public key $PK$ which is published, and the system master key $MK$ that is kept private.
    \item $AttrKeyGen(PK,MK,L)\rightarrow{SK}_L$: The attribute key generation algorithm is run by AA. The system public key $PK$, the system master key $MK$, and an attribute list $L$ are inputs of this algorithm. It returns the attribute secret key ${SK}_L$ associated with the attribute list $L$ as output.
    \item $DOParamSetup\left(PK\right)\rightarrow(PP,SP)$: The data owner parameters setup algorithm is run by DO. The system public key $PK$ is taken as input and the outputs of the algorithm are the data public parameter $PP$ and the data secret parameter $SP$.
    \item $RKeyGen(PP)\rightarrow RK$: The re-encryption key generation algorithm is run by DO to obtain a proxy re-encryption key. The data public parameter $PP$ is taken as input and the re-encryption key $RK$ is returned as the output.

    \item $OfflineEncrypt(PK,PP)\rightarrow{CT}_{off}$: The offline encryption algorithm is run by Dev while it is offline. This algorithm takes the system public key $PK$ and the data public parameter $PP$ as input. It outputs an offline ciphertext ${CT}_{off}$.

    \item $OnlineEncrypt(M,{CT}_{off})\rightarrow{CT}_{M,cloud}$: The online encryption algorithm is run by Dev while it is online. This algorithm takes some message $M$, and an offline ciphertext ${CT}_{off}$ as input. It outputs a message ciphertext ${CT}_{M,cloud}$.

    \item $AnonEncrypt(PK,PP,SP,RK,A)\rightarrow{CT}_{T_0,cloud}^A$: The anonymous encryption algorithm is run by DO. This algorithm takes the system public key $PK$, the data public parameter $PP$, the data secret parameter $SP$, the re-encryption key $RK$, and an access policy $A=\bigvee_{j=1}^{m}W_j$ as inputs. It outputs a policy ciphertext ${CT}_{T_0,cloud}^A=\{{CT}_{W_j,T_0},{\widetilde{CT}}_{W_j,T_0}\}_{1\le j\le m}$, where for ${1\le j\le m}$, the two components ${CT}_{W_j,T_0}$ and ${\widetilde{CT}}_{W_j,T_0}$ are associated with $W_j$.

    \item $Reencrypt(PP,RK,T_l,{CT}_{cloud}^A)\rightarrow{CT}_{M,T_l,user}^A$: The re-encryption algorithm is run by CSP. The data public parameter $PP$, the re-encryption key $RK$, a timestamp $T_l$, and some cloud ciphertext ${CT}_{cloud}^A=({CT}_{M,cloud},{CT}_{T_0,cloud}^A=\{{CT}_{W_j,T_0},{\widetilde{CT}}_{W_j,T_0}\}_{1\le j\le m})$ are the inputs of the algorithm. The output is a user ciphertext under the hidden access policy $A$ with respect to the timestamp $T_l$, denoted as ${CT}_{M,T_l,user}^A=({CT}_{M,T_l,user}, {CT}_{T_l,user}^A=\{{CT}_{W_j,T_l},\ {\widetilde{CT}}_{W_j,T_l}\}_{1\le j\le m})$. Here, ${CT}_{M,T_l,user}$ is the re-encrypted version of ${CT}_{M,cloud}$, and for ${1\le j\le m}$, ${CT}_{W_j,T_l}$ and ${\widetilde{CT}}_{W_j,T_l}$ are the re-encrypted versions of ${CT}_{W_j,T_0}$ and ${\widetilde{CT}}_{W_j,T_0}$ respectively.

    \item $AnonDecrypt(PK,PP,{CT}_{M,T_l,user}^A,{SK}_L)\rightarrow M$ or $\bot$: The anonymous decryption algorithm is run by DU. The system public key $PK$, the data public parameter $PP$, some user ciphertext ${CT}_{M,T_l,user}^A=({CT}_{M,T_l,user},{CT}_{T_l,user}^A=\{{CT}_{W_j,T_l},\ {\widetilde{CT}}_{W_j,T_l}\}_{1\le j\le m})$, and the attribute secret key ${SK}_L$ are inputs of the algorithm. The output is the original message $M$ or $\bot$. This algorithm consists of two phases: \emph{matching phase} and \emph{decryption phase}.

\begin{enumerate}
    \item \emph{Matching phase:} If $L\nvDash W_j$ for all $1\le j\le m$, this phase returns $\bot$ and anonymous decryption algorithm is terminated. Otherwise, the subsequent decryption phase is run.
    \item \emph{Decryption phase:} This phase returns the message $M$.
\end{enumerate}

\end{enumerate}

\begin{figure}[t]
    \centering
        \parbox{1\linewidth}{
            \begin{subfigure}[b]{1\linewidth}
            \includegraphics[width=\linewidth]{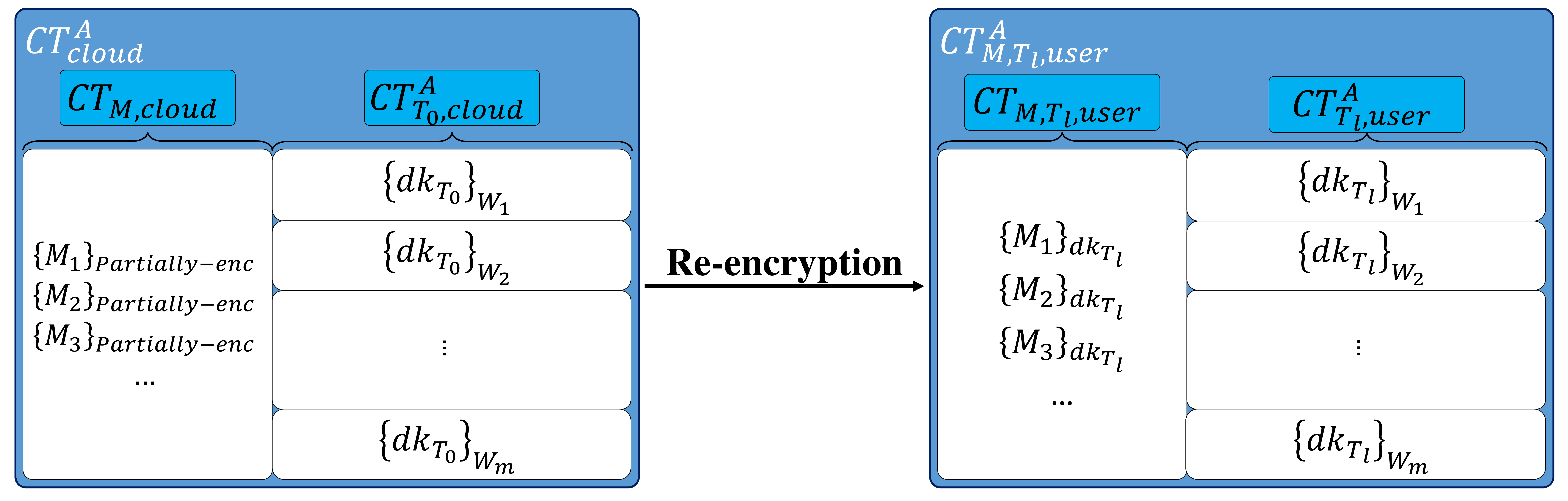}
             \caption{Re-encryption.}
             \label{fig:Reencryption}
            \end{subfigure}

            \begin{subfigure}[b]{1\linewidth}
            \includegraphics[width=\linewidth]{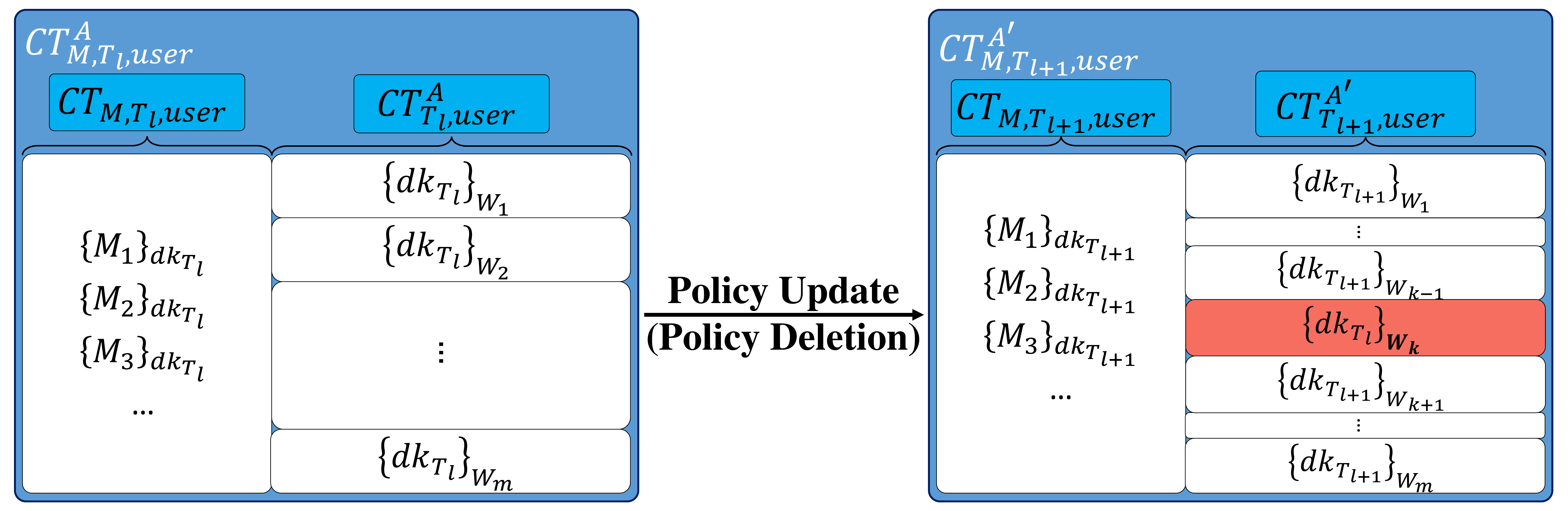}
            \caption{Policy Deletion.}
             \label{fig:Policy_Deletion}
            \end{subfigure}

            \begin{subfigure}[b]{1\linewidth}
            \includegraphics[width=\linewidth]{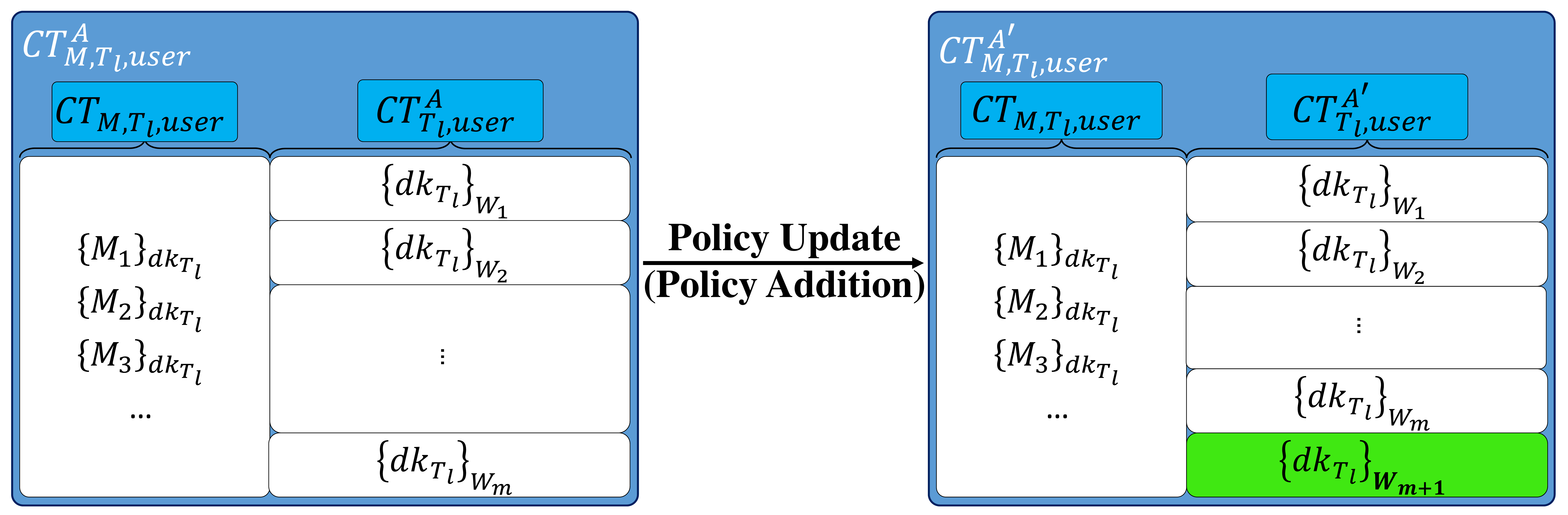}
            \caption{Policy addition.}
             \label{fig:Policy_Addition}
            \end{subfigure}
        }

    \caption{Ciphertexts.}
\end{figure}

Now, an overview of the HUAP scheme is given in the following:
\begin{enumerate}[label=\arabic*)]
    \item \emph{System initialization:} The AA runs the $SystemSetup$ algorithm to generate the system public key $PK$ and the system master key $MK$. $PK$ is published by AA and $MK$ is kept private to itself. Then, it runs $AttrKeyGen$ algorithm when it receives a request from an authorized DU. The generated attribute secret key is returned to the DU.

    \item \emph{Data initialization:} The DO runs $DOParamSetup$ algorithm to generate the data public parameter $PP$ and the data secret parameter $SP$. He publishes $PP$ and keeps $SP$ confidential. DO also runs $RKeyGen$ algorithm and sends the generated re-encryption key $RK$ to the CSP through a secure channel.

    \item \emph{Data outsource:} Before the message to be determined, Dev uses the system public key $PK$ and the data public parameter $PP$ to prepare some offline ciphertexts ${CT}_{off}$ by running $OfflineEncrypt$ algorithm. Once a message $M$ is known, Dev runs $OnlineEncrypt$ algorithm to encrypt the message and calculate the message ciphertext ${CT}_{M,cloud}$. Then, Dev sends ${CT}_{M,cloud}$ to the CSP.
    On the other side, the DO uses the data secret parameter $SP$ to calculate a data decryption key ${dk}_{T_0}$. Then, she/he defines a hidden access policy $A=\bigvee_{j=1}^{m}W_j$ and encrypts the data decryption key ${dk}_{T_0}$ under $A$ by running $AnonEncrypt$ algorithm. The output is a policy ciphertext ${CT}_{T_0,cloud}^A=\{{CT}_{W_j,T_0},{\widetilde{CT}}_{W_j,T_0}\}_{1\le j\le m}$, in which for ${1\le j\le m}$, ${CT}_{W_j,T_0}$ represents the data decryption key $dk_{T_0}$ encrypted under sub-policy $W_j$, and ${\widetilde{CT}}_{W_j,T_0}$ is a blind access policy. Afterward, the DO sends ${CT}_{T_0,cloud}^A$ to the CSP.
    When the CSP receives ${CT}_{cloud}^A=({CT}_{M,cloud},{CT}_{T_0,cloud}^A=\{{CT}_{W_j,T_0},{\widetilde{CT}}_{W_j,T_0}\}_{1\le j\le m})$ from Dev and DO, it runs $Reencrypt$ algorithm using the current timestamp $T_l$ to generate a user ciphertext ${CT}_{M,T_l,user}^A=({CT}_{M,T_l,user},{CT}_{T_l,user}^A=\{{CT}_{W_j,T_l},{\widetilde{CT}}_{W_j,T_l}\}_{1\le j\le m})$. Here, for ${1\le j\le m}$, the component ${CT}_{W_j,T_l}$ represents the data decryption key $dk_{T_l}$ encrypted under sub-policy $W_j$. The data decryption key $dk_{T_l}$ can be used to retrieve the underlying message $M$ from ${CT}_{M,T_l,user}$. Finally, ${CT}_{M,T_l,user}^A$ is published by the CSP for DUs. Fig. \ref{fig:Reencryption} shows the cloud ciphertext ${CT}_{cloud}^A$ and the user ciphertext ${CT}_{M,T_l,user}^A$ during re-encryption.

    \item \emph{Data access:} When a DU with attribute secret key ${SK}_L$ wants to decrypt a user ciphertext ${CT}_{M,T_l,user}^A$ which is encrypted under a hidden access policy $A=\bigvee_{j=1}^{m}W_j$, for $1\le j\le m$, she/he runs the $matching$ phase of $AnonDecrypt$ algorithm to check whether his attribute secret key ${SK}_L$ satisfies $W_j$ or not. If $L\models W_j$ for some $j$, DU runs the $decryption$ phase, and obtains the associated message $M$. Otherwise, $AnonDecrypt$ algorithm returns $\bot$.

    \item \emph{Access policy update:} A DO can update his defined access policy $A=\bigvee_{j=1}^{m}W_j$ at any time. Let ${CT}_{cloud}^A=({CT}_{M,cloud},{CT}_{T_0,cloud}^A=\{{CT}_{W_j,T_0},{\widetilde{CT}}_{W_j,T_0}\}_{1\le j\le m})$ be a cloud ciphertext associated with a hidden access policy $A$. Assume that the CSP has re-encrypted the cloud ciphertext ${CT}_{cloud}^A$ using the timestamp $T_l$, and has published the resulted user ciphertext ${CT}_{M,T_l,user}^A=({CT}_{M,T_l,user},{CT}_{T_l,user}^A=\{{CT}_{W_j,T_l},\ {\widetilde{CT}}_{W_j,T_l}\}_{1\le j\le m})$. Moreover, assume that the DO wants to update the access policy $A$ by defining a new access policy, $A^\prime$. In the following, the policy deletion and the policy addition operations are described. It can be shown that all possible updates of an access policy can be made by performing combinations of policy deletion and policy addition operations.

\begin{enumerate}
            \item \emph{Policy deletion:} If a DO wants to revoke some DUs whose attributes satisfy $W_k$, $1\le k\le m$, he should define a new access policy as $A^\prime=\bigvee_{j=1,j\neq k}^{m}W_j$. So, if  DO is online, he only should send a request to the CSP. Otherwise, he should determine an expiration date for $W_k$. Upon receiving the request, or reaching the expiration date, the CSP deletes the components ${CT}_{W_k,T_0}$ and ${\widetilde{CT}}_{W_k,T_0}$ from the cloud ciphertext ${CT}_{cloud}^A$. Then, DO re-encrypts the new cloud ciphertext ${CT}_{cloud}^{A^\prime}$ by running $Reencrypt$ algorithm. The output of the algorithm is a user ciphertext ${CT}_{M,T_{l+1},user}^{A^\prime}$ which is under the new access policy ${A^\prime}$ and the new timestamp $T_{l+1}$. Fig. \ref{fig:Policy_Deletion} shows a user ciphertext published by the CSP for DUs during policy deletion operation.
            \item \emph{Policy addition:} If a DO wants to expand the access policy $A$, he should define a new access policy as $A^\prime=\left(\bigvee_{j=1}^{m}W_j\right)\bigvee W_{m+1}$. Therefore, DO just should provide two new components ${CT}_{W_{m+1},T_0}$ and ${\widetilde{CT}}_{W_{m+1},T_0}$ associated with $W_{m+1}$. Then, by re-encrypting these two components, the CSP computes ${CT}_{W_{m+1},T_l}$ and ${\widetilde{CT}}_{W_{m+1},T_l}$. Then, the CSP appends the re-encrypted components to the user ciphertext ${CT}_{T_l,user}^A$. Hence, there is no need for the CSP to change the previously published components ${CT}_{M,T_l,user}$ or ${CT}_{T_l,user}^A$. Fig. \ref{fig:Policy_Addition} shows a user ciphertext generated by the CSP for DUs during policy addition operation.
\end{enumerate}

\end{enumerate}

\subsection{Security Model}

AA is assumed to be trusted. The CSP is assumed honest but curious. It executes the given protocol correctly, but it may try to obtain additional information about the stored data. All DUs are assumed to be malicious; they try to learn some unauthorized information about data stored in the CSP.
Also, it is assumed that the CSP does not collude with DUs, while unauthorized DUs may collude with each other to access the data outsourced to the CSP.

\subsection{Design Goals}
\label{section:design_goals}

The following security and performance goals are considered in our proposed scheme.

\begin{enumerate}[label=\arabic*)]
    \item \emph{Access policy update:} The DO should be able to update defined access policies. In particular, the DO should be able to revoke the access right of a group of DUs at any time, whether she/he is online or even offline.

    \item \emph{Fine-grained access control:} The DO should be able to define a desired access policy for each part of his data.

    \item \emph{Data confidentiality:} The CSP and unauthorized DUs must not be able to access the outsourced data. 

    \item \emph{Collusion resistance:} Multiple malicious data users may collude with each other to access some stored data by combining their attribute secret keys. Our scheme must resist such collusion attacks.

    \item \emph{Attribute privacy protection:}
    In many applications such as s-health, the access policy itself is considered as sensitive information and must be hidden.
    Therefore, the CSP and unauthorized DUs must not be able to obtain any information about the access policies defined by the DO.

    \item \emph{Cost efficiency:} The computational cost on DUs, DOs, and Devs should be as low as possible.

\end{enumerate}

\section{HUAP: ATTRIBUTE-BASED ACCESS CONTROL SUPPORTING HIDDEN UPDATABLE ACCESS POLICIES}

	In this section, we present our anonymous CP-ABE scheme that supports hidden updatable access policies.
	The proposed scheme utilizes online/offline encryption to reduce the computational cost for resource-constrained devices.
	Our scheme also enables data owners to outsource a major part of the access policy update process to the CSP without the need to generate new re-encryption keys. Our proposed outsourcing approach is based on blind access policies.
	In fact, the data owner defines a blind access policy and sends it to the cloud along with other ciphertext components.
	This blind access policy enables the cloud to share random parameters with authorized data users determined by the data owner.
	Whenever an access policy update is required, the cloud itself generates a new random parameter and re-encrypts all the past and future encrypted messages based on this random parameter such that the revoked data users cannot decrypt these re-encrypted messages. Then the cloud utilizes a blind access policy to share this random parameter with other authorized data users. In particular, the cloud cannot obtain any information about the identities or the attributes of these authorized data users.

	In this scheme, we split the ciphertext into two major parts. The first part is associated with encrypted messages and is generated accumulatively by resource-constrained devices. The second part is corresponding to the hidden access policy which is defined by the data owner. In fact, the messages are encrypted in the first part and the corresponding decryption key is encrypted under a hidden access policy in the second part. Therefore, a data owner can define hidden access policies while the devices that generate and encrypt messages do not need to know anything about the defined access policies.

\subsection{Our Proposed Construction}
\begin{enumerate}[label=\arabic*)]
    \item \textbf{\emph{$SystemSetup(1^\lambda)$:}} AA chooses two cyclic multiplicative groups $\mathbb{G}$ and $\mathbb{G}_T$ of a large prime order $p$, $g$ as a generator of $\mathbb{G}$ and $e:\mathbb{G}\times\mathbb{G}\rightarrow\mathbb{G}_T$ as a bilinear pairing. Let $H:\left\{0,1\right\}^\ast\rightarrow\mathbb{G}$, $\widehat{H}:\left\{0,1\right\}^\ast\rightarrow Z_p^\ast$ be two hash functions, and $F:\mathbb{G}_T\rightarrow\mathbb{G}$ be a function mapping elements of $\mathbb{G}_T$ to elements of $\mathbb{G}$.
        AA also chooses uniformly at random $y\in_RZ_p$ and $g_1,g_2,g_3,g_4\in_R\mathbb{G}$, and computes $Y={e(g_1,g_2)}^y$. The system public key is published as $PK=(g,g_1,g_2,g_3,g_4,Y)$ and the system master key $MK=(y)$ is kept private by AA.

    \item \textbf{\emph{$DOParamSetup(PK)$:}} DO chooses ${mk}_0,{mk}_1,sk\in_RZ_p^\ast$, uniformly  at random. Then the data public parameter $PP=(Q_0=g_3^{sk}$,${PP}_0={e\left(g_3,g_4\right)}^{{mk}_0}$,${PP}_1=g_3^{{mk}_1})$ is published and the data secret parameter $SP=({mk}_0,{mk}_1,{SK}_1=g_4^{{mk}_0},sk)$ is kept private by DO.

    \item \textbf{\emph{$AttrKeyGen(PK,MK,L)$:}} Assume that AA wants to generate an attribute secret key corresponding to an attribute list $L=\left[L_1,L_2,\ldots,L_n\right]$. 
    Also, assume that the universal attribute set is $\mathbb{U}=\left\{\omega_1,\omega_2,\ldots,\omega_n\right\}$ and each attribute supports multiple values, where the multi-value set for $\omega_i$ is $S_i=\left\{v_{i,1},v_{i,2},\ldots,v_{i,n_i}\right\}$. 
    For $1\le i\le n$, AA chooses $r_i\in_RZ_p$ such that $\sum_{i=1}^{n}r_i=y$. Also, AA chooses ${\hat{r}}_i\in_RZ_p$ for $1\le i\le n$ and computes $\hat{r}=\sum_{i=1}^{n}{\hat{r}}_i$. Then AA chooses $r,\lambda,\hat{\lambda}\in_RZ_p$ and computes $D_0=g_2^\lambda$, ${\widehat{D}}_0=g_1^{\hat{\lambda}}$, $D_{\mathrm{\Delta},0}=g_1^r$ and ${\widehat{D}}_{\mathrm{\Delta},0}=g_2^{y-\hat{r}}$. For $1\le i\le n$, suppose that $L_i=v_{i,k_i}$, AA computes $\left[D_{\mathrm{\Delta},i},D_{i,1},{\widehat{D}}_{i,1}\right]$ as follows:
        \begin{flalign*}
            &\left[
            \begin{matrix}
            \begin{split}
                &D_{\mathrm{\Delta},i}=g_2^{{\hat{r}}_i}.{H(i||v_{i,k_i})}^r,
                D_{i,1}=g_1^{r_i}.{H\left(0\left|\left|i\right|\right|v_{i,k_i}\right)}^\lambda,\\
                &{\widehat{D}}_{i,1}=g_2^{r_i}.{H\left(1\left|\left|i\right|\right|v_{i,k_i}\right)}^{\hat{\lambda}}
            \end{split}
            \end{matrix}
            \right].&
        \end{flalign*}

        \noindent Finally, the attribute secret key is ${SK}_L=\langle D_0,{\widehat{D}}_0,D_{\mathrm{\Delta},0},{\widehat{D}}_{\mathrm{\Delta},0}, \left\{D_{\mathrm{\Delta},i},D_{i,1},{\widehat{D}}_{i,1}\right\}_{1\le i\le n}\rangle$.

    \item \textbf{\emph{$RKeyGen(PP)$:}} DO selects $s_{cloud}\in_RZ_p$ and sets re-encryption key as $RK=s_{cloud}$. Then DO sends the re-encryption key $RK$ to the CSP through a secure channel. Therefore, the DO and the CSP will be able to compute $S_{T_l}=\widehat{H}(s_{cloud}||l)$ for $l\in\mathbb{Z}^+$.

    \item \textbf{\emph{$OfflineEncrypt(PK,PP)$:}} Dev chooses $r_d\in_RZ_p^\ast$ and computes offline ciphertext ${CT}_{off}$ as follows:
        \begin{flalign*}
            &{CT}_{off}=\left(
            \begin{matrix}
            \begin{split}
                &U_0=g_3^{r_d},
                U_1=g_3^{r_d.sk},
                V_0=e\left(g_3,g_4\right)^{r_d.{mk}_0}
            \end{split}
            \end{matrix}
            \right).&
        \end{flalign*}

    \item \textbf{\emph{$OnlineEncrypt(M,{CT}_{off})$:}} Once the message $M$ is determined, Dev calculates the message ciphertext ${CT}_{M,cloud}$, signs it and finally sends it to the CSP:
        \begin{flalign*}
            &{CT}_{M,cloud}=\left(
                U_0, U_1, V=M.V_0
            \right).&
        \end{flalign*}

    \item \textbf{\emph{$AnonEncrypt(PK,PP,SP,RK,A)$:}} Suppose that a DO wants to define an access policy $A=\bigvee_{j=1}^{m}W_j$ where $W_j=\left[W_{j,1},W_{j,2},\ldots,W_{j,n}\right]$. First, DO computes the data decryption key with respect to the timestamp $T_0$ as ${dk}_{T_0}=g_4^{{mk}_0}.g_3^{sk.{mk}_1}.g_3^{{mk}_1.S_{T_0}}$. Then, for $1\le j\le m$:
        DO chooses $s_{1,j},s_{1,j}^\prime,s_{1,j}^{\prime\prime},s_{2,j},s_{2,j}^{\prime\prime}\in_RZ_p^\ast$. Also, for $1\le i\le n$, the DO chooses $\{\sigma_{i,j,\Delta},\sigma_{i,j,0},\sigma_{i,j,1},\sigma_{i,j,0}^\prime,\sigma_{i,j,1}^\prime\in_R\mathbb{G} |1\le i\le n\}$ such that $\prod_{i=1}^{n}\sigma_{i,j,\Delta}=\prod_{i=1}^{n}\sigma_{i,j,0}=\prod_{i=1}^{n}\sigma_{i,j,1}=\prod_{i=1}^{n}\sigma_{i,j,0}^\prime=\prod_{i=1}^{n}\sigma_{i,j,1}^\prime=1_\mathbb{G}$. Then, the DO computes $[C_{i,t,\Delta,w_j}, C_{i,t,0,w_j},{\widehat{C}}_{i,t,0,w_j},C_{i,t,0,w_j}^\prime,{\widehat{C}}_{i,t,0,w_j}^\prime]$ for $1\le i\le n$ as follows:
        \begin{itemize}
          \item[$$] \begin{enumerate}
              \item If $v_{i,t}\in W_{j,i}$, then
                \begin{flalign*}
                    &\left[
                    \begin{matrix}
                    \begin{split}
                        C_{i,t,\mathrm{\Delta},w_j}&=\sigma_{i,j,\mathrm{\Delta}}.{H(i||v_{i,t})}^{s_{1,j}^\prime},\\
                        C_{i,t,0,w_j}&=\sigma_{i,j,0}.{H\left(0\left|\left|i\right|\right|v_{i,t}\right)}^{s_{1,j}^{\prime\prime}},\\
                        {\widehat{C}}_{i,t,0,w_j}&=\sigma_{i,j,1}.{H\left(1\left|\left|i\right|\right|v_{i,t}\right)}^{{s_{1,j}-s}_{1,j}^{\prime\prime}},\\
                        C_{i,t,0,w_j}^\prime&=\sigma_{i,j,0}^\prime.{H\left(0\left|\left|i\right|\right|v_{i,t}\right)}^{s_{2,j}^{\prime\prime}},\\
                        {\widehat{C}}_{i,t,0,w_j}^\prime&=\sigma_{i,j,1}^\prime.{H(1||i||v_{i,t})}^{{s_{2,j}-s}_{2,j}^{\prime\prime}}
                    \end{split}
                    \end{matrix}
                    \right].&
                \end{flalign*}

              \item If $v_{i,t}\notin W_{j,i}$, then $[C_{i,t,\mathrm{\Delta},w_j},C_{i,t,0,w_j},{\widehat{C}}_{i,t,0,w_j},C_{i,t,0,w_j}^\prime,{\widehat{C}}_{i,t,0,w_j}^\prime]$ are random elements in $\mathbb{G}$.

            \end{enumerate}
        \end{itemize}

        \noindent Then, DO computes ${CT}_{W_j,T_0}$ and ${\widetilde{CT}}_{W_j,T_0}$ as follows:
        \begin{flalign*}
            &{\widetilde{CT}}_{W_j,T_0}=&\\
            &\left(
            \begin{matrix}
            \begin{split}
                &{\widetilde{C}}_{w_j}^\prime=Y^{s_{2,j}},
                C_{1,w_j}^\prime=g_2^{s_{2,j}^{\prime\prime}},
                {\widehat{C}}_{1,w_j}^\prime=g_1^{s_{2,j}-s_{2,j}^{\prime\prime}},\\
                &\{\{
                    \begin{matrix}
                    \begin{split}
                        &C_{i,t,0,w_j}^\prime,{\widehat{C}}_{i,t,0,w_j}^\prime
                    \end{split}
                    \end{matrix}
                \}_{1\le t\le n_i}\}_{1\le i\le n}
            \end{split}
            \end{matrix}
            \right),&
        \end{flalign*}

        \begin{flalign*}
            &{CT}_{W_j,T_0}=&\\
            &\left(
            \begin{matrix}
            \begin{split}
                &{\widetilde{C}}_{w_j}={dk}_{T_0}.F\left(Y^{s_{1,j}}\right),
                C_{\Delta,w_j}=Y^{s_{1,j}^\prime},\\
                &{\widehat{C}}_{0,w_j}=g_1^{s_{1,j}^\prime},
                C_{1,w_j}=g_2^{s_{1,j}^{\prime\prime}},
                {\widehat{C}}_{1,w_j}=g_1^{s_{1,j}-s_{1,j}^{\prime\prime}},\\
                &\{\{
                    \begin{matrix}
                    \begin{split}
                        &C_{i,t,\Delta,w_j},
                        C_{i,t,0,w_j},
                        {\widehat{C}}_{i,t,0,w_j}
                    \end{split}
                    \end{matrix}
                \}_{1\le t\le n_i}\}_{1\le i\le n}
            \end{split}
            \end{matrix}
            \right).&
        \end{flalign*}

        \noindent Finally, the policy ciphertext which is prepared for sending to the cloud is as follows, where ${\widetilde{CT}}_{W_j,T_0}$ is a blind access policy for $1\le j\le m$:
        \begin{flalign*}
            &{CT}_{T_0,cloud}^A=
            \begin{matrix}
            \begin{split}
                \{{CT}_{W_j,T_0},{\widetilde{CT}}_{W_j,T_0}\}_{1\le j\le m}
            \end{split}
            \end{matrix}
            .&
        \end{flalign*}

    \item \textbf{\emph{$Reencrypt(PP,RK,T_l,{CT}_{cloud}^A)$:}} Suppose that the CSP wants to re-encrypt the ciphertext ${CT}_{cloud}^A=({CT}_{M,cloud},{CT}_{T_0,cloud}^A=\{{CT}_{W_j,T_0},{\widetilde{CT}}_{W_j,T_0}\}_{1\le j\le m})$ for the timestamp $T_l$, where the underlying access policy is $A=\bigvee_{j=1}^{m}W_j$. First, the CSP chooses $r^\prime\in_RZ_p^\ast$, calculates $S_{T_l}=\widehat{H}(RK||l)$, and computes ${CT}_{M,T_l,user}$ as follows:
        \begin{flalign*}
            &{CT}_{M,T_l,user}=\left[
            \begin{matrix}
            \begin{split}
                U_0^{T_l}&=U_0.g_3^{r^\prime},\\
                U_1^{T_l}&=U_1.g_3^{r^\prime.sk}.\left(U_0^{T_l}\right)^{S_{T_l}},\\
                V^{T_l}&=V.{e\left(g_3,g_4\right)}^{{mk}_0.r^\prime}
            \end{split}
            \end{matrix}
            \right].&
        \end{flalign*}

        Then, for $1\le j\le m$, suppose that ${CT}_{W_j,T_0}$ and ${\widetilde{CT}}_{W_j,T_0}$ are as the same as in $AnonEncrypt$ algotithm. CSP chooses $r_{w_j}^{\prime\prime},r_{w_j}\in_RZ_p^\ast$ and computes ${\widetilde{CT}}_{W_j,T_l}$ and ${CT}_{W_j,T_l}$ as follows:
        \begin{flalign*}
            &{\widetilde{CT}}_{W_j,T_l}=&\\
            &\left(
            \begin{matrix}
            \begin{split}
                &{\widetilde{C}}_{w_j,T_l}^\prime=Y^{r_{w_j}}.\left({\widetilde{C}}_{w_j}^\prime\right)^{r_{w_j}^{\prime\prime}},\\
                &C_{1,{w_j,T}_l}^\prime=\left(C_{1,w_j}^\prime\right)^{r_{w_j}^{\prime\prime}},
                {\widehat{C}}_{1,{w_j,T}_l}^\prime=\left({\widehat{C}}_{1,w_j}^\prime\right)^{r_{w_j}^{\prime\prime}},\\
                &\left\{\left\{
                    \begin{matrix}
                    \begin{split}
                        &C_{i,t,0,{w_j,T}_l}^\prime=\left(C_{i,t,0,w_j}^\prime\right)^{r_{w_j}^{\prime\prime}},\\
                        &{\widehat{C}}_{i,t,0,{w_j,T}_l}^\prime=\left({\widehat{C}}_{i,t,0,w_j}^\prime\right)^{r_{w_j}^{\prime\prime}}
                    \end{split}
                    \end{matrix}
                \right\}_{1\le t\le n_i}\right\}_{1\le i\le n}
            \end{split}
            \end{matrix}
            \right),&
        \end{flalign*}

        \begin{flalign*}
            &{CT}_{W_j,T_l}=&\\
            &\left(
            \begin{matrix}
            \begin{split}
                &{\widetilde{C}}_{{w_j,T}_l}=F\left(Y^{r_{w_j}}\right).g_3^{{mk}_1.\left(S_{T_l}-S_{T_0}\right)}.{\widetilde{C}}_{w_j},\\
                &{C}_{\Delta,w_j,T_l}={C}_{\Delta,w_j},
                {\widehat{C}}_{0,{w_j,T}_l}={\widehat{C}}_{0,w_j},\\
                &C_{1,{w_j,T}_l}=C_{1,w_j},
                {\widehat{C}}_{1,{w_j,T}_l}={\widehat{C}}_{1,w_j},\\
                &\left\{\left\{
                    \begin{matrix}
                    \begin{split}
                        &{C}_{i,t,\Delta,w_j,T_l}={C}_{i,t,\Delta,w_j},\\
                        &C_{i,t,0,w_j,T_l}=C_{i,t,0,w_j},\\
                        &{\widehat{C}}_{i,t,0,{w_j,T}_l}={\widehat{C}}_{i,t,0,w_j}
                    \end{split}
                    \end{matrix}
                \right\}_{1\le t\le n_i}\right\}_{1\le i\le n}
            \end{split}
            \end{matrix}
            \right).&
        \end{flalign*}
        \noindent where ${\widetilde{C}}_{{w_j,T}_l}=F\left(Y^{r_{w_j}}\right).{dk}_{T_l}.F\left(Y^{s_{1,j}}\right)$. Here, for $1\le j\le m$, ${\widetilde{CT}}_{W_j,T_0}$ is a blind access policy that is utilized to share the randomly generated parameter ${Y}^{r_{w_j}}$ with authorized data users.
        Finally, the re-encrypted ciphertext which is prepared for data users with respect to the timestamp $T_l$ is:
        \begin{flalign*}
            &{CT}_{M,T_l,user}^A=&\\
            &\left(
            \begin{matrix}
            \begin{split}
                &{CT}_{M,T_l,user},\\
                &{CT}_{T_l,user}^A=\left\{
                        {CT}_{W_j,T_l},{\widetilde{CT}}_{W_j,T_l}
                \right\}_{1\le j\le m}
            \end{split}
            \end{matrix}
            \right).&
        \end{flalign*}

    \item \emph{$AnonDecrypt(PK,PP,{CT}_{M,T_l,user},{CT}_{W_j,T_l},{\widetilde{CT}}_{W_j,T_l},\allowbreak{SK}_L)$:} DU tests and decrypts ciphertext ${CT}_{M,T_l,user}$ with attribute secret key ${SK}_L$ in two following phases:
        \begin{enumerate}
                  \item \emph{matching phase:} For $1\le i\le n$, suppose that $L_i=v_{i,t}$. $L\models{W}_j$ if and only if the following equation holds:
                    \begin{flalign} \label{equation(1)}
                        &{C}_{\Delta,w_j,T_l}=\frac{e\left({\widehat{C}}_{0,w_j,T_l}, {\widehat{D}}_{\Delta,0}.\prod_{i=1}^{n}{D}_{\Delta,i}\right)}{e\left(\prod_{i=1}^{n}{C}_{i,t,\Delta,w_j,T_l},{D}_{\Delta,0}\right)}.&
                    \end{flalign}

                    \noindent If $L\models{W}_j$, the subsequent decryption phase is started. Otherwise, the algorithm $AnonDecrypt$ returns $\bot$.

                  \item \emph{decryption phase:} Suppose that $L\models{W}_j$ and $L_i=v_{i,t}$ for $1\le i\le n$. At first, the DU computes $Y^{r_{w_j}}$ and $Y^{s_{1,j}}$ as follows:
                    \begin{flalign} \label{equation(2)}
                        Y^{r_{w_j}}={\widetilde{C}}_{{w_j,T}_l}^\prime
                        &.\frac{
                                e\left(\prod_{i=1}^{n}C_{i,t,0,{w_j,T}_l}^\prime,D_0\right)
                            }{
                                e\left(C_{1,w_j,T_l}^\prime,\prod_{i=1}^{n}D_{i,1}\right)
                            }& \notag\\
                            &.\frac
                            {
                                e\left(\prod_{i=1}^{n}{\widehat{C}}_{i,t,0,w_j,T_l}^\prime,{\widehat{D}}_0\right)
                            }
                            {
                                e\left({\widehat{C}}_{1,w_j,T_l}^\prime,\prod_{i=1}^{n}{\widehat{D}}_{i,1}\right)
                            },&
                    \end{flalign}

                    \begin{flalign} \label{equation(3)}
                        Y^{s_{1,j}}=
                        &\frac{
                                e\left(C_{1,w_j,T_l},\prod_{i=1}^{n}D_{i,1}\right)
                            }{
                                e\left(\prod_{i=1}^{n}C_{i,t,0,{w_j,T}_l},D_0\right)
                            }& \notag\\
                            .&\frac
                            {
                                e\left({\widehat{C}}_{1,w_j,T_l},\prod_{i=1}^{n}{\widehat{D}}_{i,1}\right)
                            }
                            {
                                e\left(\prod_{i=1}^{n}{\widehat{C}}_{i,t,0,w_j,T_l},{\widehat{D}}_0\right)
                            }.&
                    \end{flalign}

                    \noindent Then, the DU computes ${dk}_{T_l}$ as follows:

                    \begin{flalign} \label{equation(4)}
                        {dk}_{T_l}=&\frac{{\widetilde{C}}_{w_j,T_l}}{F\left(Y^{r_{w_j}}\right).F\left(Y^{s_{1,j}}\right)}.&\\
                        \notag
                    \end{flalign}

                    \noindent Finally, the DU retrieves the message $M$ as follows:

                    \begin{flalign} \label{equation(5)}
                        M=&\sfrac{V^{T_l}}{\frac{e\left(U_0^{T_l},{dk}_{T_l}\right)}{e\left({PP}_1,U_1^{T_l}\right)}}.&\\
                        \notag
                    \end{flalign}
        \end{enumerate}
\end{enumerate}

\subsection{Consistency of the Proposed Construction}

In the following, we show the correctness of Equations (\ref{equation(1)}) to (\ref{equation(5)}).
Firstly, the attributes satisfy the access policy if and only if (\ref{equation(1)}) holds as it is shown in the following:
\begin{flalign*}
    &\frac{
            e\left({\widehat{C}}_{0,w_j,T_l},{\widehat{D}}_{\Delta,0}.\prod_{i=1}^{n}{D}_{\Delta,i}\right)
        }{
            e\left(\prod_{i=1}^{n}{C}_{i,t,\Delta,w_j,T_l},{D}_{\Delta,0}\right)
        }&\\
        &=\ \frac
        {
            e\left(g_1^{s_{1,j}^\prime},g_2^{y-\hat{r}}.\prod_{i=1}^{n}{g_2^{{\hat{r}}_i}.{H(i||v_{i,k_i})}^r}\right)
        }
        {
            e\left(\prod_{i=1}^{n}{\sigma_{i,j,\mathrm{\Delta}}.{H(i||v_{i,t})}^{s_{1,j}^\prime}},g_1^r\right)
        }&\\
        &=\ \frac
        {
            e\left(g_1^{s_{1,j}^\prime},g_2^y.\prod_{i=1}^{n}{H(i||v_{i,k_i})}^r\right)
        }
        {
            e\left(\prod_{i=1}^{n}{H(i||v_{i,t})}^{s_{1,j}^\prime},g_1^r\right)
        }&\\
        &=\ e\left(g_1^{s_{1,j}^\prime},g_2^y\right)=\left({e\left(g_1,g_2\right)}^y\right)^{s_{1,j}^\prime}={C}_{\Delta,w_j,T_l}.&
\end{flalign*}

\noindent The correctness of (\ref{equation(2)}) is shown in the following:
\begin{flalign*}
	&{\widetilde{C}}_{{w_j,T}_l}^\prime.
    \frac
    {
        e\left(\prod_{i=1}^{n}C_{i,t,0,{w_j,T}_l}^\prime,D_0\right).e\left(\prod_{i=1}^{n}{\widehat{C}}_{i,t,0,w_j,T_l}^\prime,{\widehat{D}}_0\right)
    }
    {
        e\left(C_{1,w_j,T_l}^\prime,\prod_{i=1}^{n}D_{i,1}\right).e\left({\widehat{C}}_{1,w_j,T_l}^\prime,\prod_{i=1}^{n}{\widehat{D}}_{i,1}\right)
    }&\\
    =&\ {\widetilde{C}}_{{w_j,T}_l}^\prime.
    \frac
    {
        e\left(\prod_{i=1}^{n}\left(C_{i,t,0,w_j}^\prime\right)^{r_{w_j}^{\prime\prime}},g_2^\lambda\right)
    }
    {
        e\left(\left(C_{1,w_j}^\prime\right)^{r_{w_j}^{\prime\prime}},\prod_{i=1}^{n}{g_1^{r_i}.{H\left(0\left|\left|i\right|\right|v_{i,k_i}\right)}^\lambda}\right)
    }&\\
    \ .&\frac
    {
        e\left(\prod_{i=1}^{n}\left({\widehat{C}}_{i,t,0,w_j}^\prime\right)^{r_{w_j}^{\prime\prime}},g_1^{\hat{\lambda}}\right)
    }
    {
        e\left(\left({\widehat{C}}_{1,w_j}^\prime\right)^{r_{w_j}^{\prime\prime}},\prod_{i=1}^{n}{g_2^{r_i}.{H\left(1\left|\left|i\right|\right|v_{i,k_i}\right)}^{\hat{\lambda}}}\right)
    }&\\
    =&\ {\widetilde{C}}_{{w_j,T}_l}^\prime.
    \frac
    {
        e\left(\prod_{i=1}^{n}\left(\sigma_{i,j,0}^\prime.{H\left(0\left|\left|i\right|\right|v_{i,t}\right)}^{s_{2,j}^{\prime\prime}}\right)^{r_{w_j}^{\prime\prime}},g_2^\lambda\right)
    }
    {
        e\left(\left(g_2^{s_{2,j}^{\prime\prime}}\right)^{r_{w_j}^{\prime\prime}},\prod_{i=1}^{n}{g_1^{r_i}.{H\left(0\left|\left|i\right|\right|v_{i,k_i}\right)}^\lambda}\right)
    }&\\
    \ .&\frac
    {
        e\left(\prod_{i=1}^{n}{(\sigma_{i,j,1}^\prime.{H(1||i||v_{i,t})}^{{s_{2,j}-s}_{2,j}^{\prime\prime}})}^{r_{w_j}^{\prime\prime}},g_1^{\hat{\lambda}}\right)
    }
    {
        e\left(\left(g_1^{s_{2,j}-s_{2,j}^{\prime\prime}}\right)^{r_{w_j}^{\prime\prime}},\prod_{i=1}^{n}{g_2^{r_i}.{H\left(1\left|\left|i\right|\right|v_{i,k_i}\right)}^{\hat{\lambda}}}\right)
    }&\\
    =&\ {\widetilde{C}}_{{w_j,T}_l}^\prime.
    \frac
    {
        e\left(\prod_{i=1}^{n}H\left(0\left|\left|i\right|\right|v_{i,t}\right),g_2^{\lambda.r_{w_j}^{\prime\prime}.s_{2,j}^{\prime\prime}}\right)
    }
    {
        e\left(g_2^{r_{w_j}^{\prime\prime}.s_{2,j}^{\prime\prime}},g_1^y.\prod_{i=1}^{n}{H\left(0\left|\left|i\right|\right|v_{i,k_i}\right)}^\lambda\right)
    }&\\
    \ .&\frac
    {
        e\left(\prod_{i=1}^{n}{H(1||i||v_{i,t})},g_1^{\hat{\lambda}.r_{w_j}^{\prime\prime}.\left({s_{2,j}-s}_{2,j}^{\prime\prime}\right)}\right)
    }
    {
        e\left(g_1^{r_{w_j}^{\prime\prime}.\left(s_{2,j}-s_{2,j}^{\prime\prime}\right)},g_2^y.\prod_{i=1}^{n}{H\left(1\left|\left|i\right|\right|v_{i,k_i}\right)}^{\hat{\lambda}}\right)
    }&\\
    =&\ {\widetilde{C}}_{{w_j,T}_l}^\prime.\frac{1}{e\left(g_1^{r_{w_j}^{\prime\prime}.s_{2,j}},g_2^y\right)}=\frac{Y^{r_{w_j}}.\left({\widetilde{C}}_{w_j}^\prime\right)^{r_{w_j}^{\prime\prime}}}{e\left(g_1^{r_{w_j}^{\prime\prime}.s_{2,j}},g_2^y\right)}&\\
    =&\ \frac{Y^{r_{w_j}}.Y^{r_{w_j}^{\prime\prime}.s_{2,j}}}{e\left(g_1^{r_{w_j}^{\prime\prime}.s_{2,j}},g_2^y\right)}=Y^{r_{w_j}}.&
\end{flalign*}

\noindent For showing the correctness of (\ref{equation(3)}) we have:
\begin{flalign*}
    &\frac
    {
        e\left(C_{1,w_j,T_l},\prod_{i=1}^{n}D_{i,1}\right).e\left({\widehat{C}}_{1,w_j,T_l},\prod_{i=1}^{n}{\widehat{D}}_{i,1}\right)
    }
    {
        e\left(\prod_{i=1}^{n}C_{i,t,0,{w_j,T}_l},D_0\right).e\left(\prod_{i=1}^{n}{\widehat{C}}_{i,t,0,w_j,T_l},{\widehat{D}}_0\right)
    }&\\
    =&\ \frac
    {
        e\left(g_2^{s_{1,j}^{\prime\prime}},\ \prod_{i=1}^{n}{g_1^{r_i}.{H\left(0\left|\left|i\right|\right|v_{i,k_i}\right)}^\lambda}\right)
    }
    {
        e\left(\prod_{i=1}^{n}{\sigma_{i,j,0}.{H\left(0\left|\left|i\right|\right|v_{i,t}\right)}^{s_{1,j}^{\prime\prime}}},g_2^\lambda\right)
    }&\\
    \ .&\frac
    {
        e\left(g_1^{s_{1,j}-s_{1,j}^{\prime\prime}},\prod_{i=1}^{n}{g_2^{r_i}.{H\left(1\left|\left|i\right|\right|v_{i,k_i}\right)}^{\hat{\lambda}}}\right)
    }
    {
        e\left(\prod_{i=1}^{n}{\sigma_{i,j,1}.{H\left(1\left|\left|i\right|\right|v_{i,t}\right)}^{{s_{1,j}-s}_{1,j}^{\prime\prime}}},g_1^{\hat{\lambda}}\right)
    }&\\
    =&\ \frac
    {
        e\left(g_2^{s_{1,j}^{\prime\prime}},g_1^y.\prod_{i=1}^{n}{H\left(0\left|\left|i\right|\right|v_{i,k_i}\right)}^\lambda\right)
    }
    {
        e\left(\prod_{i=1}^{n}{H\left(0\left|\left|i\right|\right|v_{i,t}\right)}^{s_{1,j}^{\prime\prime}},g_2^\lambda\right)
    }&\\
    \ .&\frac
    {
        e\left(g_1^{s_{1,j}-s_{1,j}^{\prime\prime}},g_2^y.\prod_{i=1}^{n}{H\left(1\left|\left|i\right|\right|v_{i,k_i}\right)}^{\hat{\lambda}}\right)
    }
    {
        e\left(\prod_{i=1}^{n}{H\left(1\left|\left|i\right|\right|v_{i,t}\right)}^{{s_{1,j}-s}_{1,j}^{\prime\prime}},g_1^{\hat{\lambda}}\right)
    }&\\
    =&\ e\left(g_1^{s_{1,j}},\ g_2^y\right)=Y^{s_{1,j}}.&
\end{flalign*}

\noindent Then, the data decryption key ${dk}_{T_l}$ is retrieved using (\ref{equation(4)}) as follows:
\begin{flalign*}
    &\frac{{\widetilde{C}}_{w_j,T_l}}{F\left(Y^{r_{w_j}}\right).F\left(Y^{s_{1,j}}\right)}=\frac{F\left(Y^{r_{w_j}}\right).{dk}_{T_l}.F\left(Y^{s_{1,j}}\right)}{F\left(Y^{r_{w_j}}\right).F\left(Y^{s_{1,j}}\right)}={dk}_{T_l}.&
\end{flalign*}

\noindent Finally, the message $M$ can be recovered using (\ref{equation(5)}) as follows:
\begin{flalign*}
    &\frac
    {
        e\left(U_0^{T_l},\ {dk}_{T_l}\right)
    }
    {
        e\left({PP}_1,\ U_1^{T_l}\right)
    }
    =\frac
    {
        e\left(g_3^{r_d+r^\prime},g_4^{{mk}_0}.g_3^{{mk}_1.sk}.g_3^{{mk}_1.S_{T_l}}\right)
    }
    {
        e\left(g_3^{{mk}_1},g_3^{\left(r_d+r^\prime\right).sk}.\left(U_0^{T_l}\right)^{S_{T_l}}\right)
    }&\\
    =&\ \frac
    {
        e\left(g_3^{r_d+r^\prime},g_4^{{mk}_0}\right).e\left(g_3^{r_d+r^\prime},g_3^{{mk}_1.sk}\right).e\left(g_3^{r_d+r^\prime},g_3^{{mk}_1.S_{T_l}}\right)
    }
    {
        e\left(g_3^{{mk}_1},g_3^{\left(r_d+r^\prime\right).sk}\right).e\left(g_3^{{mk}_1},\left(g_3^{r_d+r^\prime}\right)^{S_{T_l}}\right)
    }&\\
    =&\ {e\left(g_3,g_4\right)}^{{mk}_0.\left(r_d+r^\prime\right)}=A,&\\
\end{flalign*}

\noindent and then,
\begin{flalign*}
    &M=\frac{V^{T_l}}{\frac{e\left(U_0^{T_l},{dk}_{T_l}\right)}{e\left({PP}_1,U_1^{T_l}\right)}}=\frac{M.e\left(g_3^{{mk}_0},g_4^{r_d+r^\prime}\right)}{A}.&
\end{flalign*}

\section{SECURITY ANALYSIS}

In this section, we prove the security of HUAP scheme in the random oracle model.

\subsection{Formal Security Model and Definition}

The security of our proposed scheme is proven in the indistinguishability against selective ciphertext-policy and chosen-plaintext attacks (IND-sCP-CPA) security model \cite{5,9,36}. The model is an interactive game between an adversary and a challenger. The adversary attempts to (1) obtain some information about a plaintext from the corresponding ciphertext, (2) distinguish the access policies embedded in ciphertexts.\\

\textbf{IND-sCP-CPA Game. }

\textbf{Init:} $\mathcal{A}$ submits two challenge access policies $A_0^\ast$ and $A_1^\ast$ to the challenger. $\mathcal{A}$ also submits a timestamp $T_l$.

\textbf{Setup:} The challenger $\mathcal{S}$ specifies a security parameter $\lambda$, and runs the $SystemSetup$ algorithm to get a system master key $MK$ and the corresponding system public key $PK$. Also, $\mathcal{S}$ runs the $DOParamSetup$ algorithm to get a data secret parameter $SP$ and the corresponding data public parameter $PP$. Moreover, $\mathcal{S}$ runs the $RKeyGen$ algorithm to get a re-encryption key $RK$. It keeps $MK$, $SP$, and $RK$ secretly and gives $PK$ and $PP$ to $\mathcal{A}$.

\textbf{Phase 1:} The adversary $\mathcal{A}$ makes some queries to the following oracles:

\begin{itemize}
        \item[$\bullet$] \textbf{AttrKeyGen oracle $\mathcal{O}_{AttrKeyGen}$:} $\mathcal{A}$ submits an attribute list $L$. The challenger runs the $AttrKeyGen$ algorithm and returns the corresponding attribute secret key ${SK}_L$ to $\mathcal{A}$ only if $L\nvDash A_0^\ast \land L\nvDash A_1^\ast$. Otherwise, it outputs $\bot$.

        \item[$\bullet$] \textbf{Reencrypt oracle $\mathcal{O}_{Reencrypt}$:} The adversary $\mathcal{A}$ submits a timestamp $T_i$, and a cloud ciphertext ${CT}_{cloud}^A$. The challenger runs the $Reencrypt$ algorithm and returns the corresponding user ciphertext ${CT}_{M,T_i,user}^A$ only if $i\le l$. Otherwise, it outputs $\bot$.

\end{itemize}

\textbf{Challenge:} When \textbf{Phase 1} is over, $\mathcal{A}$ sends two different equal-length messages $M_0$ and $M_1$ to the challenger. Afterwards, the challenger first runs $OfflineEncrypt$ algorithm to get an offline ciphertext $CT_{off}$. Then, the challenger randomly selects a bit $\nu\in\left\{0,1\right\}$, and computes the message ciphertext ${CT}_{M_\nu,cloud}=OnlineEncrypt(M_\nu,CT_{off})$. The challenger also computes the policy ciphertext $CT_{T_0,cloud}^{{A_\nu^\ast}}=AnonEncrypt(PK,PP,SP,RK,{A_\nu^\ast})$. Finally, the challenger returns the cloud ciphertext ${CT}_{cloud}^{A_\nu^\ast}=\{{CT}_{M_\nu,cloud},CT_{T_0,cloud}^{{A_\nu^\ast}}\}$ to $\mathcal{A}$.

\textbf{Phase 2:} It is similar to \textbf{Phase 1}.

\textbf{Guess:} $\mathcal{A}$ outputs a bit $\nu^\prime\in\left\{0,\ 1\right\}$ as a guess of $\nu$, and it wins the game if $\nu^\prime=\nu$.\\
In the IND-sCP-CPA game, we define the advantage of $\mathcal{A}$ as follows:\\
${\rm Adv}_{HUAP}^{IND-sCP-CPA}\left(\mathcal{A}\right)=\left|Pr\left[\nu^\prime=\ \nu\right]-\frac{1}{2}\right|$.\\

\textbf{Definition 1.} HUAP is said to be IND-sCP-CPA secure if the advantage of a probabilistic polynomial-time adversary to win the IND-sCP-CPA game is a negligible function in the security parameter.

\subsection{Security Proof}

In the following, we prove that our proposed scheme is secure according to \textbf{Definition 1}.
This security proof extends the security proof presented in\cite{5}.\\

\textbf{Theorem 1.} The HUAP scheme is IND-sCP-CPA secure in the random oracle model such that $\epsilon_{CPA}\le2\epsilon_{DBDH}+n.\epsilon_{DL}$, where $\epsilon_{CPA}$ denotes the advantage of a polynomial-time adversary $\mathcal{A}$ to win the IND-sCP-CPA game, $\epsilon_{DBDH}$ and $\epsilon_{DL}$ respectively denote the advantage of a distinguisher of a DBDH challenge and a D-Linear challenge, and $n$ represents the total number of attributes in the universe.

\textbf{Proof.} 
The proof of this theorem is provided through three lemmas. In this proof, we marginally alter the original game $G$ into a sequence of hybrid games denoted by $\left\{G_0^\prime,G_0,G_1,\ldots,G_{h_{max}}\right\}$. 
Firstly, in \textbf{Lemma 1}, to establish the first hybrid game $G_0^\prime$, we embed a DBDH challenge into the ciphertext by substituting the challenge ciphertext component $V$ with a random element in $\mathbb{G}_T$, while the other components are generated in a routine manner. 
In this fashion, we construct a distinguisher of DBDH challenge with the help of the distinguisher of $G$ and $G_0^\prime$. 
Subsequently, in \textbf{Lemma 2}, to form the next hybrid game $G_0$, we embed another DBDH challenge into the ciphertext by substituting the challenge ciphertext component ${\widetilde{C}}_{w_{\nu}^{\ast}}$ with a random element in $\mathbb{G}$.
Afterward, in \textbf{Lemma 3}, we establish a series of hybrid games denoted by $G_h$, where $1\le h\le h_{max}$, by embedding D-Linear challenges into the corresponding ciphertexts as follows:
Suppose that $h_{max}$ is the number of attribute values like $v_{i,t}$ satisfying $(v_{i,t}\in W_{0,i}^\ast\land v_{i,t}\notin W_{1,i}^\ast)$ or $(v_{i,t}\notin W_{0,i}^\ast\land v_{i,t}\in W_{1,i}^\ast)$. For each of these attribute values, we modify the game $G_{h-1}$ into a game $G_h$ by substituting the ciphertext components $\{\{C_{i,t,\Delta,w_\nu^\ast}, C_{i,t,0,w_\nu^\ast},{\widehat{C}}_{i,t,0,w_\nu^\ast},C_{i,t,0,w_\nu^\ast}^\prime,{\widehat{C}}_{i,t,0,w_\nu^\ast}^\prime\}_{1\le t\le n_i}\}_{1\le i\le n}$ with random elements, while other components are generated normally. 
The process is repeated until there is no attribute value $v_{i,t}$ satisfying $(v_{i,t}\in W_{0,i}^\ast\land v_{i,t}\notin W_{1,i}^\ast)$ or $(v_{i,t}\notin W_{0,i}^\ast\land v_{i,t}\in W_{1,i}^\ast)$. 
In the last game denoted by $G_{h_{max}}$, the challenge ciphertext components are chosen independently from the random bit $\nu$, and hence the adversary $\mathcal{A}$ has not any advantage in winning the game.

Through \textbf{Lemma 1 }to \textbf{Lemma 3}, we prove that $|Pr\left[\mathcal{E}\right]-Pr\left[\mathcal{E}_0^\prime\right]|\le\epsilon_{DBDH}$, $\left|Pr\left[\mathcal{E}_0^\prime\right]-Pr\left[\mathcal{E}_0\right]\right|\le\epsilon_{DBDH}$, and $\left|Pr\left[\mathcal{E}_{h-1}\right]-Pr\left[\mathcal{E}_h\right]\right|\le\epsilon_{DL}$ for $1\le h\le h_{max}$, where $\mathcal{E}$, $\mathcal{E}_0^\prime$, and $\mathcal{E}_h$ respectively denote the event that $\mathcal{A}$ wins the game $G$, $G_0^\prime$, and $G_h$. 
Thus, $\epsilon_{CPA}=\left|Pr\left[\mathcal{E}\right]-\frac{1}{2}\right|=\left|Pr\left[\mathcal{E}\right]-Pr\left[\mathcal{E}_{h_{max}}\right]\right|$, and from the triangle inequality, we have:
\begin{flalign*}
	\epsilon_{CPA}\le
	&\ \left|Pr\left[\mathcal{E}\right]-Pr\left[\mathcal{E}_0^\prime\right]\right|+\left|Pr\left[\mathcal{E}_0^\prime\right]-Pr\left[\mathcal{E}_0\right]\right|&\\
	&+\sum_{h=1}^{h_{max}}\left|Pr\left[\mathcal{E}_{h-1}\right]-Pr\left[\mathcal{E}_h\right]\right|.&
\end{flalign*}

\noindent Therefore, it is obviously concluded that $\epsilon_{CPA}\le2\epsilon_{DBDH}+n.\epsilon_{DL}$.\\

\textbf{Lemma 1.} Under the DBDH assumption, the difference between advantages of $\mathcal{A}$ in games $G$ and $G_0^\prime$ is negligible such that $\left|Pr\left[\mathcal{E}\right]-Pr\left[\mathcal{E}_0^\prime\right]\right|\le\epsilon_{DBDH}$.

\textbf{Proof.} We show that if $\epsilon_0^\prime=\left|Pr\left[\mathcal{E}\right]-Pr\left[\mathcal{E}_0^\prime\right]\right|$ is not negligible, then there exists a simulator $\mathcal{S}_0^\prime$ that can break the DBDH assumption. To construct $\mathcal{S}_0^\prime$, suppose that it is given a DBDH instance $[g,A,B,C,Z] = [g,g^a,g^b,g^c,Z]$ by the challenger where $g,A,B,C\in\mathbb{G}$ and $Z\in\mathbb{G}_T$, and at the other side, it plays the role of a challenger for the adversary $\mathcal{A}$. In this manner, $\mathcal{S}_0^\prime$ will be able to win the DBDH game with the non-negligible advantage $\epsilon_0^\prime$ by exploiting $\mathcal{A}$. Accordingly, the simulator $\mathcal{S}_0^\prime$ acts as follows:

\comment{
Suppose that $\epsilon_0^\prime$ is the difference between the advantages of the adversary $\mathcal{A}$ in the games $G$ and $G_0^\prime$. We build a simulator $\mathcal{S}_0^\prime$ that can play the DBDH game with advantage $\epsilon_0^\prime$. Given a DBDH challenge tuple $[g,A,B,C,Z] = [g,g^a,g^b,g^c,Z]$ by the challenger where $Z$ is either ${e\left(g,g\right)}^{abc}$ or a random element in $\mathbb{G}_T$ with equal probability, the simulation proceeds as follows:
}

\textbf{Init:} The adversary $\mathcal{A}$ gives $\mathcal{S}_0^\prime$ two challenge access policies $A_0^\ast=W_0^\ast=\left[W_{0,1}^\ast,\ldots,W_{0,n}^\ast\right]$ and $A_1^\ast=W_1^\ast=\left[W_{1,1}^\ast,\ldots,W_{1,n}^\ast\right]$. Afterward, $\mathcal{S}_0^\prime$ selects a random bit $\nu\in\left\{0,1\right\}$.

\textbf{Setup:} The simulator $\mathcal{S}_0^\prime$ selects $g_1,g_2\in_R\mathbb{G}$, and $\omega,y,sk,{mk}_1\in_RZ_p^\ast$, sets $g_3=g^\omega$, $g_4=B$, and computes $Y={e(g_1,g_2)}^y$, $Q_0=g_3^{sk}$, ${PP}_1=g_3^{{mk}_1}$, ${PP}_0={e\left(g_3,\ g_4\right)}^{{mk}_0}={e\left(g^{{mk}_0},\ B\right)}^\omega={e\left(A,\ B\right)}^\omega$, which implies that ${mk}_0=a$. 
Afterwards, $\mathcal{S}_0^\prime$ sends the system public key $PK=\langle g,g_1,g_2,g_3,g_4,Y \rangle$ and the data public parameter $PP=(Q_0,{PP}_0,{PP}_1)$ to $\mathcal{A}$.

\textbf{Phase 1:} The simulator $\mathcal{S}_0^\prime$ answers  $\mathcal{A}$'s queries by simulating $\mathcal{O}_{Hash}$ and $\mathcal{O}_{AttrKeyGen}$ as follows:

\begin{itemize}
        \item[$\bullet$] \textbf{Hash Query $\mathcal{O}_{Hash}\left(\mathcal{M}\right)$:} Firstly, the simulator $\mathcal{S}_0^\prime$ selects an empty list $\mathcal{L}_H$. When the adversary $\mathcal{A}$ queries the random oracle $\mathcal{O}_{Hash}\left(.\right)$ for an input $\mathcal{M}$, $\mathcal{S}_0^\prime$ checks in the list $\mathcal{L}_H$ if the value of $H(\mathcal{M})$ has been defined. If there exists a record in $\mathcal{L}_H$ associated to the queried point $\mathcal{M}$, $\mathcal{S}_0^\prime$ returns the previously defined value. 
Otherwise, $\mathcal{S}_0^\prime$ selects $\{\{\tau_{i,t},a_{i,t},b_{i,t}\in_RZ_p\}_{1\le t\le n_i}\}_{1\le i\le n}$, and calculates the output as follows:
            \begin{enumerate}[label=\arabic*)]
              \item For $\mathcal{M}=(i||v_{i,t})$, $\mathcal{S}_0^\prime$ sets $\alpha=\tau_{i,t}$, and returns $H(\mathcal{M})=g^\alpha$.
              \item For $\mathcal{M}=(0||i||v_{i,t})$, $\mathcal{S}_0^\prime$ sets $\alpha=a_{i,t}$, and returns $H(\mathcal{M})=g^\alpha$.
              \item For $\mathcal{M}=(1||i||v_{i,t})$, $\mathcal{S}_0^\prime$ sets $\alpha=b_{i,t}$, and returns $H(\mathcal{M})=g^\alpha$.
            \end{enumerate}
        \noindent Finally, $\mathcal{S}_0^\prime$ adds $\langle \mathcal{M},\alpha,H(\mathcal{M}) \rangle$ to the list $\mathcal{L}_H$.

        \item[$\bullet$] \textbf{AttrKeyGen Query $\mathcal{O}_{AttrKeyGen}\left(L\right)$:} When the adversary $\mathcal{A}$ queries the random oracle $\mathcal{O}_{AttrKeyGen}\left(.\right)$ for an attribute list $L=[L_1,L_2,\ldots,L_n]$ where $L_i=v_{i,k_i}$ and with the restriction that $(L\nvDash A_0^\ast\land L\nvDash A_1^\ast)$, the simulator $\mathcal{S}_0^\prime$ firstly selects $r_1,r_2,\ldots,r_{n}\in_RZ_p$ such that $\sum_{i=1}^{n}r_i=y$.
Also, $\mathcal{S}_0^\prime$ selects $r,\lambda,\hat{\lambda}\in_RZ_p$ and $\left\{{\hat{r}}_i\in_RZ_p\right\}_{1\le i\le n}$, sets $\hat{r}=\sum_{i=1}^{n}{\hat{r}}_i$, and computes $D_0=g_2^\lambda$, ${\widehat{D}}_0=g_1^{\hat{\lambda}}$, $D_{\mathrm{\Delta},0}=g_1^r$, ${\widehat{D}}_{\mathrm{\Delta},0}=g_2^{y-\hat{r}}$. 
Afterwards, $\mathcal{S}_0^\prime$ computes the rest of the attribute secret key components for $1\le i\le n$:
\begin{flalign*} \label{secret_key}
                &\left[
                \begin{matrix}
                \begin{split}
                &D_{\mathrm{\Delta},i}=g_2^{{\hat{r}}_i}.{H(i||v_{i,k_i})}^r=g_2^{{\hat{r}}_i}.{g^{r.\tau_{i,k_i}}},\\
                &D_{i,1}=g_1^{r_i}.{H\left(0||i||v_{i,k_i}\right)}^\lambda=g_1^{r_i}.{g^{\lambda.a_{i,k_i}}},\\
                &{\widehat{D}}_{i,1}=g_2^{r_i}.{H\left(1||i||v_{i,k_i}\right)}^{\hat{\lambda}}=g_2^{r_i}.{g^{\hat{\lambda}.b_{i,k_i}}}
                \end{split}
                \end{matrix}
                \right].&
\end{flalign*}

\noindent Finally, the following attribute secret key is returned:
${SK}_L=\langle D_0, {\widehat{D}}_0, D_{\mathrm{\Delta},0}, {\widehat{D}}_{\mathrm{\Delta},0}, {\{ D_{\mathrm{\Delta},i}, D_{i,1}, {\widehat{D}}_{i,1} \}}_{1\le i\le n} \rangle$.

\end{itemize}

\textbf{Challenge:} The adversary $\mathcal{A}$ sends two different equal-length messages $M_0$ and $M_1$ to the simulator $\mathcal{S}_0^\prime$. This latter sets $V=M_\nu.Z^\omega$ and hence when ${Z=e\left(g,g\right)}^{abc}$, we have $V=M_\nu$.${e\left(g,g\right)}^{abc\omega}=M_\nu.{e\left(g_3,g_4\right)}^{ac}=M_\nu.{e\left(g_3,g_4\right)}^{{mk}_0.c}=M_\nu.{PP}_0^{r_d}$, which implies that $r_d=c$. 
Afterwards, $\mathcal{S}_0^\prime$ calculates $U_0=g_3^{r_d}=g^{r_d.\omega}=\left(g^{r_d}\right)^\omega=\left(g^c\right)^\omega$, $U_1=g_3^{r_d.sk}=g^{r_d.sk.\omega}=\left(g^{r_d}\right)^{sk.\omega}=\left(g^c\right)^{sk.\omega}$, ${dk}_{T_0}=g_4^{{mk}_0}.g_3^{sk.{mk}_1}.g_3^{{mk}_1.S_{T_0}}=g_3^{sk.{mk}_1}.g^{\omega.{mk}_1.S_{T_0}+b.a}=g_3^{sk.{mk}_1}.g^\alpha$ where $\alpha\in_RZ_p^\ast$, and sets $S_{T_0}=\sfrac{\left(\alpha-b.a\right)}{\left(\omega.{mk}_1\right)}$.
Moreover, $\mathcal{S}_0^\prime$ selects $s_1,s_1^\prime,s_1^{\prime\prime},s_2,s_2^{\prime\prime}\in_RZ_p^\ast$ and computes ${\widetilde{C}}_{w_\nu^\ast}^\prime=1.Y^{s_2}$, $C_{1,w_\nu^\ast}^\prime=g_2^{s_2^{\prime\prime}}$, ${\widehat{C}}_{1,w_\nu^\ast}^\prime=g_1^{s_2-s_2^{\prime\prime}}$, ${\widetilde{C}}_{w_\nu^\ast}={dk}_{T_0}.F\left(Y^{s_1}\right)$, $C_{\Delta,w_\nu^\ast}=Y^{s_1^\prime}$, ${\widehat{C}}_{0,w_\nu^\ast}=g_1^{s_1^\prime}$, $C_{1,w_\nu^\ast}=g_2^{s_1^{\prime\prime}}$, ${\widehat{C}}_{1,w_\nu^\ast}=g_1^{s_1-s_1^{\prime\prime}}$. 
Then, $\mathcal{S}_0^\prime$ picks $\{\sigma_{i,\Delta},\sigma_{i,0},\sigma_{i,1},\sigma_{i,0}^\prime,\sigma_{i,1}^\prime\in_RG|1\le i\le n\}$ such that $\prod_{i=1}^{n}\sigma_{i,\Delta}=\prod_{i=1}^{n}\sigma_{i,0}=\prod_{i=1}^{n}\sigma_{i,1}=\prod_{i=1}^{n}\sigma_{i,0}^\prime=\prod_{i=1}^{n}\sigma_{i,1}^\prime=1_\mathbb{G}$, and computes $[C_{i,t,\Delta,w_\nu^\ast},C_{i,t,0,w_\nu^\ast},{\widehat{C}}_{i,t,0,w_\nu^\ast},C_{i,t,0,w_\nu^\ast}^\prime,{\widehat{C}}_{i,t,0,w_\nu^\ast}^\prime]$ as follows:

\begin{enumerate}
      \item If $v_{i,t}\in W_{\nu,i}^\ast$, then
        \begin{flalign*}
            &\left[
            \begin{matrix}
            \begin{split}
                C_{i,t,\mathrm{\Delta},w_\nu^\ast}&=\sigma_{i,\mathrm{\Delta}}.H(i||v_{i,t})^{s_{1}^\prime}=\sigma_{i,\mathrm{\Delta}}.g^{\tau_{i,t}.s_{1}^\prime},\\
                C_{i,t,0,w_\nu^\ast}&=\sigma_{i,0}.H(0||i||v_{i,t})^{s_{1}^{\prime\prime}}=\sigma_{i,0}.g^{a_{i,t}.s_{1}^{\prime\prime}},\\
                {\widehat{C}}_{i,t,0,w_\nu^\ast}&=\sigma_{i,1}.H(1||i||v_{i,t})^{({s_{1}-s}_{1}^{\prime\prime})}\\&=\sigma_{i,1}.g^{b_{i,t}.({s_{1}-s}_{1}^{\prime\prime})},\\
                C_{i,t,0,w_\nu^\ast}^\prime&=\sigma_{i,0}^\prime.H(0||i||v_{i,t})^{s_{2}^{\prime\prime}}=\sigma_{i,0}^\prime.g^{a_{i,t}.s_{2}^{\prime\prime}},\\
                {\widehat{C}}_{i,t,0,w_\nu^\ast}^\prime&=\sigma_{i,1}^\prime.H(1||i||v_{i,t})^{{(s_{2}-s}_{2}^{\prime\prime})}\\&=\sigma_{i,1}^\prime.g^{b_{i,t}.{(s_{2}-s}_{2}^{\prime\prime})}
            \end{split}
            \end{matrix}
            \right].&
        \end{flalign*}
      \item If $v_{i,t}\notin W_{\nu,i}^\ast$, then $[C_{i,t,\mathrm{\Delta},w_\nu^\ast},C_{i,t,0,w_\nu^\ast},{\widehat{C}}_{i,t,0,w_\nu^\ast},C_{i,t,0,w_\nu^\ast}^\prime,{\widehat{C}}_{i,t,0,w_\nu^\ast}^\prime]$ are chosen randomly from $\mathbb{G}$.
\end{enumerate}

\noindent Finally, the simulator $\mathcal{S}_0^\prime$ returns the following challenge ciphertext of $M_\nu$ with respect to $A_\nu^\ast$:
\begin{flalign*}
    &{CT}_{cloud}^{A_\nu^\ast}=&\\
    &\left(
    \begin{matrix}
    \begin{split}
    	&U_0,U_1,V,
        {\widetilde{C}}_{w_\nu^\ast}^\prime,C_{1,w_\nu^\ast}^\prime,{\widehat{C}}_{1,w_\nu^\ast}^\prime,\\
        &\{\{
            \begin{matrix}
            \begin{split}
                &C_{i,t,0,w_\nu^\ast}^{\prime},{\widehat{C}}_{i,t,0,w_\nu^\ast}^{\prime}
            \end{split}
            \end{matrix}
        \}_{1\le t\le n_i}\}_{1\le i\le n},\\
        &{\widetilde{C}}_{w_\nu^\ast},C_{\Delta,w_\nu^\ast},{\widehat{C}}_{0,w_\nu^\ast},C_{1,w_\nu^\ast},{\widehat{C}}_{1,w_\nu^\ast},\\
        &\{\{
            \begin{matrix}
            \begin{split}
                &C_{i,t,\Delta,w_\nu^\ast},C_{i,t,0,w_\nu^\ast},{\widehat{C}}_{i,t,0,w_\nu^\ast}
            \end{split}
            \end{matrix}
        \}_{1\le t\le n_i}\}_{1\le i\le n}
    \end{split}
    \end{matrix}
    \right).&
\end{flalign*}

\textbf{Phase 2:} The adversary $\mathcal{A}$ continues to query the oracles as in \textbf{Phase 1}.

\textbf{Guess:} The adversary $\mathcal{A}$ guesses a bit $\nu^\prime$ as the value of $\nu$. $\mathcal{A}$ wins the game if $\nu^\prime=\nu$. When $\mathcal{A}$ wins the game, $\mathcal{S}_0^\prime$ returns $1$ and otherwise returns $0$ to the DBDH challenger.
Indicate that if ${Z=e\left(g,g\right)}^{abc}$, then $\mathcal{A}$ is in game $G$, and otherwise, $Z$ is a random element in $\mathbb{G}_T$ and $\mathcal{A}$ is in game $G_0^\prime$.
Therefore the advantage of $\mathcal{S}_0^\prime$ in the DBDH game is equal to $\epsilon_0^\prime$, where $\epsilon_0^\prime$ is the difference between the advantages of $\mathcal{A}$ to win the game $G$ and the game $G_0^\prime$.
Finally, with respect to the DBDH assumption we have $\left|Pr\left[\mathcal{E}\right]-Pr\left[\mathcal{E}_0^\prime\right]\right|=\epsilon_0^\prime\le\epsilon_{DBDH}$.\\

\textbf{Lemma 2.} Under the DBDH assumption, the difference between advantages of $\mathcal{A}$ in games $G_0^\prime$ and $G_0$ is negligible such that $\left|Pr\left[\mathcal{E}_0^\prime\right]-Pr\left[\mathcal{E}_0\right]\right|\le\epsilon_{DBDH}$.

\textbf{Proof.} We show that if $\epsilon_0=\left|Pr\left[\mathcal{E}_0^\prime\right]-Pr\left[\mathcal{E}_0\right]\right|$ is not negligible, then there exists a simulator $\mathcal{S}_0$ that can break the DBDH assumption. To construct $\mathcal{S}_0$, suppose that it is given a DBDH instance $[g,A,B,C,Z] = [g,g^a,g^b,g^c,Z]$ by the challenger where $g,A,B,C\in\mathbb{G}$ and $Z\in\mathbb{G}_T$, and at the other side, it plays the role of a challenger for the adversary $\mathcal{A}$. In this manner, $\mathcal{S}_0$ will be able to win the DBDH game with the non-negligible advantage $\epsilon_0$ by exploiting $\mathcal{A}$. 
Accordingly, the simulator $\mathcal{S}_0$ acts as follows:

\comment{
Suppose that $\epsilon_0$ is the difference between the advantages of the adversary $\mathcal{A}$ in the games $G_0^\prime$ and $G_0$. We build a simulator $\mathcal{S}_0$ that can play the DBDH game with advantage $\epsilon_0$. Given a DBDH challenge $[g,A,B,C,Z] = [g,g^a,g^b,g^c,Z]$ by the challenger where $Z$ is either ${e\left(g,g\right)}^{abc}$ or a random element in $\mathbb{G}_T$ with equal probability, the simulation proceeds as follows:
}

\textbf{Init:} The adversary $\mathcal{A}$ gives $\mathcal{S}_0$ two challenge access policies $A_0^\ast=W_0^\ast=\left[W_{0,1}^\ast,\ldots,W_{0,n}^\ast\right]$ and $A_1^\ast=W_1^\ast=\left[W_{1,1}^\ast,\ldots,W_{1,n}^\ast\right]$, and a timestamp $T_l$. Afterward, $\mathcal{S}_0$ selects a random bit $\nu\in\left\{0,1\right\}$.

\textbf{Setup:} The simulator $\mathcal{S}_0$ selects $RK,S_{T_0},\omega,{mk}_0,{mk}_1,sk\in_RZ_p$, and $g_3,g_4\in_R\mathbb{G}$, sets $g_1=g^\omega$, $g_2=B$ and computes $Q_0=g_3^{sk}$, ${PP}_0={e\left(g_3,g_4\right)}^{{mk}_0}$, ${PP}_1=g_3^{{mk}_1}$, $Y={e\left(g_1,g_2\right)}^y={e\left(g^y,B\right)}^\omega={e\left(A,B\right)}^\omega$, which implies that $y=a$. 
Afterwards, $\mathcal{S}_0$ sends the system public key $PK=\langle g,g_1,g_2,g_3,g_4,Y \rangle$ and the data public parameter $PP=(Q_0,{PP}_0,{PP}_1)$ to $\mathcal{A}$.

\textbf{Phase 1:} The simulator $\mathcal{S}_0$ answers  $\mathcal{A}$'s queries by simulating $\mathcal{O}_{Hash}$ and $\mathcal{O}_{AttrKeyGen}$ as follows:

\begin{itemize}
        \item[$\bullet$] \textbf{Hash Query $\mathcal{O}_{Hash}\left(\mathcal{M}\right)$:} Firstly, the simulator $\mathcal{S}_0$ selects an empty list $\mathcal{L}_H$. When the adversary $\mathcal{A}$ queries the random oracle $\mathcal{O}_{Hash}\left(.\right)$ for an input $\mathcal{M}$, $\mathcal{S}_0$ checks in the list $\mathcal{L}_H$ if the value of $H(\mathcal{M})$ has been defined. If there exists a record in $\mathcal{L}_H$ associated to the queried point $\mathcal{M}$, $\mathcal{S}_0$ returns the previously defined value. Otherwise, $\mathcal{S}_0$ selects $\{\{\tau_{i,t},a_{i,t},b_{i,t}\in_RZ_p\}_{1\le t\le n_i}\}_{1\le i\le n}$, and calculates the output as follows:
            \begin{enumerate}[label=\arabic*)]
              \item For $\mathcal{M}=(i||v_{i,t}$), $\mathcal{S}_0$ sets $\alpha=\tau_{i,t}$, returns $H(\mathcal{M})=g^\alpha$ if $v_{i,t}\in W_{\nu,i}^\ast$, or $H(\mathcal{M})=g_2^\alpha$ if $v_{i,t}\notin W_{\nu,i}^\ast$.
              \item For $\mathcal{M}=(0||i||v_{i,t})$, $\mathcal{S}_0$ sets $\alpha=a_{i,t}$, and returns $H(\mathcal{M})=g_2^\alpha$.
              \item For $\mathcal{M}=(1||i||v_{i,t})$, $\mathcal{S}_0$ sets $\alpha=b_{i,t}$, returns $H(\mathcal{M})=g^\alpha$ if $v_{i,t}\in W_{\nu,i}^\ast$, or $H(\mathcal{M})=g_2^\alpha$ if $v_{i,t}\notin W_{\nu,i}^\ast$.
            \end{enumerate}

        \noindent Finally, $\mathcal{S}_0$ adds $\langle \mathcal{M},\alpha,H(\mathcal{M}) \rangle$ to the list $\mathcal{L}_H$.

        \item[$\bullet$] \textbf{AttrKeyGen Query $\mathcal{O}_{AttrKeyGen}\left(L\right)$:} Suppose that the adversary $\mathcal{A}$ queries the random oracle $\mathcal{O}_{AttrKeyGen}\left(.\right)$ for an attribute list $L=[L_1,L_2,\ldots,L_n]$ where $L_i=v_{i,k_i}$ and with the restriction that $(L\nvDash A_0^\ast\land L\nvDash A_1^\ast)$. In such a case, there must be an integer $j\in\left\{1,2,\ldots,n\right\}$ such that $L_j\notin W_{\nu,j}^\ast$. $\mathcal{S}_0$ firstly picks $\{{\hat{r}}_i^\prime\in_R Z_p\}_{1\le i\le n}$, sets ${\hat{r}}_j=a+{\hat{r}}_j^\prime$, and for $1\le i\le n$ sets ${\hat{r}}_i={\hat{r}}_i^\prime$ if $i\neq j$. 
Afterwards, $\mathcal{S}_0$ sets $\hat{r}=\sum_{i=1}^{n}{\hat{r}}_i=a+\sum_{i=1}^{n}{\hat{r}}_i^\prime$ and computes ${\widehat{D}}_{\mathrm{\Delta},0}=\prod_{i=1}^{n}g_2^{-{\hat{r}}_i^\prime}=g_2^{-\sum_{i=1}^{n}{\hat{r}}_i^\prime}=g_2^{a-\hat{r}}=g_2^{y-\hat{r}}$.
Furthermore, $\mathcal{S}_0$ selects $r_i\in_RZ_p$ for $i\neq j$ and sets $r_j=a-\sum_{i=1,i\neq j}^{n}r_i\ mod\ p$. 
Then, $\mathcal{S}_0$ computes ${\{D_{\mathrm{\Delta},i},D_{i,1},{\widehat{D}}_{i,1}\}}_{1\le i\le n}$ as follows:
            \begin{enumerate}
              \item If $i=j$, then $\mathcal{S}_0$ selects $\beta,\lambda,{\hat{\lambda}}^\prime\in_RZ_p$ and computes the following components:
                \begin{flalign*}
                &\begin{cases}
                    D_{\mathrm{\Delta},j}&=g_2^{{\hat{r}}_j}.H\left(j||v_{j,k_j}\right)^r=g_2^{{\hat{r}}_j^\prime+a}.\left(g_2^{\tau_{j,k_j}}\right)^r\\
                    &=g_2^{{\hat{r}}_j^\prime+a+r\tau_{j,k_j}}=g_2^{{\hat{r}}_j^\prime+\beta},\\
                    D_{j,1}&=g_1^{r_j}.H\left(0||j||v_{j,k_j}\right)^\lambda
                    =g_1^{a-\sum_{i=1,i\neq j}^{n}r_i}.g_2^{a_{j,k_j}\lambda}\\
                    &=\left(g^a\right)^\omega.g_2^{a_{j,k_j}\lambda}.\prod_{i=1,i\neq j}^{n}g_1^{-r_i},\\
                    {\widehat{D}}_{j,1}&=g_2^{r_j}.H\left(1||j||v_{j,k_j}\right)^{\hat{\lambda}}=g_2^{r_j}.g_2^{\hat{\lambda}b_{j,k_j}}\\
                    &=g_2^{r_j+b_{j,k_j}.\left({\hat{\lambda}}^{'}-\frac{a}{b_{j,k_j}}\right)}=g_2^{r_j-a+{\hat{\lambda}}^{'}b_{j,k_j}}\\
                    &=g_2^{{\hat{\lambda}}^{'}.b_{j,k_j}}.\prod_{i=1,i\neq j}^{n}g_2^{-r_i}\\
                \end{cases}&
                \end{flalign*}

                \noindent where $r=\sfrac{\left(\beta-a\right)}{\tau_{j,k_j}}$ and $\hat{\lambda}={\hat{\lambda}}^\prime-\left(\sfrac{a}{b_{j,k_j}}\right)$.

              \item If $i\neq j$, then $\mathcal{S}_0$ computes the following components:
                \begin{flalign*}
                &\begin{cases}
                    D_{\mathrm{\Delta},i}&=g_2^{{\hat{r}}_i}.H\left(i||v_{i,k_i}\right)^r=g_2^{{\hat{r}}_i^\prime}.\left(g^{\tau_{i,k_i}}\right)^r\\
                    &=g_2^{{\hat{r}}_i^\prime}.\left(g^\frac{\beta-a}{\tau_{j,k_j}}\right)^{\tau_{i,k_i}}\\
                    &=g_2^{{\hat{r}}_i^\prime}.g^\frac{\beta\tau_{i,k_i}}{\tau_{j,k_j}}.\left(g^a\right)^{-\frac{\tau_{i,k_i}}{\tau_{j,k_j}}},\\
                    D_{i,1}&=g_1^{r_i}.H\left(0\left|\left|i\right|\right|v_{i,k_i}\right)^\lambda=g_1^{r_i}.g_2^{a_{i,k_i}\lambda},\\
                    {\widehat{D}}_{i,1}&=g_2^{r_i}.H\left(1||i||v_{i,k_i}\right)^{\hat{\lambda}}=g_2^{r_i}.g^{\hat{\lambda}b_{i,k_i}}\\
                    &=g_2^{r_i}.g^{b_{i,k_i}{\hat{\lambda}}^\prime}.\left(g^a\right)^{-\frac{b_{i,k_i}}{b_{j,k_j}}}
                \end{cases}&
                \end{flalign*}

            \end{enumerate}

\noindent Subsequently, $\mathcal{S}_0$ computes the rest of the attribute secret key components as follows:
                \begin{flalign*}
                &\left[
                \begin{matrix}
                \begin{split}
    	            &D_0=g_2^\lambda,{\widehat{D}}_0=g_1^{\hat{\lambda}}=g_1^{{\hat{\lambda}}^\prime-\left(\sfrac{a}{b_{j,k_j}}\right)}=g_1^{{\hat{\lambda}}^\prime}.\left(g^a\right)^{-\frac{\omega}{b_{j,k_j}}},\\
    	            &D_{\mathrm{\Delta},0}=g_1^r=g_1^{\frac{\beta-a}{\tau_{j,k_j}}}=g_1^{\frac{\beta}{\tau_{j,k_j}}}.\left(g^a\right)^{-\frac{\omega}{\tau_{j,k_j}}}\\
				\end{split}
				\end{matrix}
				\right].&
                \end{flalign*}

\noindent Finally, the following attribute secret key is returned:
${SK}_L=\langle D_0, {\widehat{D}}_0, D_{\mathrm{\Delta},0}, {\widehat{D}}_{\mathrm{\Delta},0}, {\{ D_{\mathrm{\Delta},i}, D_{i,1}, {\widehat{D}}_{i,1} \}}_{1\le i\le n} \rangle$.

        \item[$\bullet$] \textbf{Reencrypt Query $\mathcal{O}_{Reencrypt}\left(T_i,{CT}_{cloud}^A\right)$:} Suppose that $\mathcal{A}$ submits a ciphertext ${CT}_{cloud}^A=({CT}_{M,cloud}=[U_0,U_1,V],{CT}_{T_0,cloud}^A=[{CT}_{W,T_0},{\widetilde{CT}}_{W,T_0}]$), and a timestamp $T_i$ where $i\le l$.\\
Firstly, $\mathcal{S}_0$ checks in the list $\mathcal{L}_{\widehat{H}}$ if the tuple $\langle i,S_{T_i}\rangle$ has been queried before. If it was, the corresponding value is retrieved. Otherwise, $\mathcal{S}_0$ chooses $S_{T_i}\in_RZ_p^\ast$ and puts the tuple $\langle i,S_{T_i}\rangle$ in the list $\mathcal{L}_{\widehat{H}}$. 
Afterwards, $\mathcal{S}_0$ chooses $r^\prime\in_RZ_p^\ast$ and computes ${CT}_{M,T_i,user}$ as follows:
            \begin{flalign*}
                &{CT}_{M,T_i,user}=
                \left[
                \begin{matrix}
                \begin{split}
                    U_0^{T_i}&=U_0.g_3^{r^\prime},\\
                    U_1^{T_i}&=U_1.g_3^{r^\prime.sk}.\left(U_0^{T_i}\right)^{S_{T_i}},\\
                    V^{T_i}&=V.e\left(g_3^{{mk}_0},g_4^{r^\prime}\right)
                \end{split}
                \end{matrix}
                \right].&
            \end{flalign*}

            \noindent Then, $\mathcal{S}_0$ chooses $r^{\prime\prime},r\in_RZ_p^\ast$ and computes ${\widetilde{CT}}_{W,T_i}$ and ${CT}_{W,T_i}$ as follows:
            \begin{flalign*}
                &{\widetilde{CT}}_{W,T_i}=&\\
                &\left(
                \begin{matrix}
                \begin{split}
                    &{\widetilde{C}}_{w,T_i}^\prime=Y^r.\left({\widetilde{C}}_w^\prime\right)^{r^{\prime\prime}},\\
                    &C_{1,{w,T}_i}^\prime=\left(C_{1,w}^\prime\right)^{r^{\prime\prime}},
                    {\widehat{C}}_{1,{w,T}_i}^\prime=\left({\widehat{C}}_{1,w}^\prime\right)^{r^{\prime\prime}},\\
                    &\left\{\left\{
                        \begin{matrix}
                        \begin{split}
                            &C_{i,t,0,{w,T}_i}^\prime=\left(C_{i,t,0,w}^\prime\right)^{r^{\prime\prime}},\\
                            &{\widehat{C}}_{i,t,0,{w,T}_i}^\prime=\left({\widehat{C}}_{i,t,0,w}^\prime\right)^{r^{\prime\prime}}
                        \end{split}
                        \end{matrix}
                    \right\}_{1\le t\le n_i}\right\}_{1\le i\le n}
                \end{split}
                \end{matrix}
                \right),&
            \end{flalign*}

            \begin{flalign*}
                &{CT}_{W,T_i}=&\\
                &\left(
                \begin{matrix}
                \begin{split}
                    &{\widetilde{C}}_{{w,T}_i}=F\left(Y^r\right).g_3^{{mk}_1.\left(S_{T_i}-S_{T_0}\right)}.{\widetilde{C}}_{w},\\
                    &{C}_{\mathrm{\Delta},w,T_i}={C}_{\mathrm{\Delta},w},
                    {\widehat{C}}_{0,{w,T}_i}={\widehat{C}}_{0,w}\\
                    &C_{1,{w,T}_i}=C_{1,w},
                    {\widehat{C}}_{1,{w,T}_i}={\widehat{C}}_{1,w},\\
                    &\left\{\left\{
                        \begin{matrix}
                        \begin{split}
                            &{C}_{i,t,\mathrm{\Delta},w,T_i}={C}_{i,t,\mathrm{\Delta},w},\\
                            &C_{i,t,0,w,T_i}=C_{i,t,0,w},\\
                            &{\widehat{C}}_{i,t,0,{w,T}_i}={\widehat{C}}_{i,t,0,w}
                        \end{split}
                        \end{matrix}
                    \right\}_{1\le t\le n_i}\right\}_{1\le i\le n}
                \end{split}
                \end{matrix}
                \right).&
            \end{flalign*}

            \noindent Finally, the following user ciphertext which is computed according to the timestamp $T_i$ is returned:
            \begin{flalign*}
                &{CT}_{M,T_i,user}^A=
                \left(
                \begin{matrix}
                \begin{split}
                    &{CT}_{M,T_i,user},\\
                    &{CT}_{T_i,user}^A=\left[{CT}_{W,T_i},{\widetilde{CT}}_{W,T_i}\right]
                \end{split}
                \end{matrix}
                \right).&
            \end{flalign*}

\end{itemize}

\textbf{Challenge:} The adversary $\mathcal{A}$ sends two different equal-length messages $M_0$ and $M_1$ to the simulator $\mathcal{S}_0$. This latter sets $V$ to be a random element in $\mathbb{G}_T$. 
Afterwards, $\mathcal{S}_0$ selects $r_d\in_RZ_p^\ast$ and sets $U_0=g_3^{r_d}$, $U_1=g_3^{r_d.sk}$, ${dk}_{T_0}=g_4^{{mk}_0}.g_3^{sk.{mk}_1}.g_3^{{mk}_1.S_{T_0}}$.
Subsequently, $\mathcal{S}_0$ calculates ${\widetilde{C}}_{w_\nu^\ast}={dk}_{T_0}.F\left(Z^\omega\right)$ and hence when $Z={e\left(g,g\right)}^{abc}$, we have ${\widetilde{C}}_{w_\nu^\ast}={dk}_{T_0}.F\left({e\left(g_1,g_2\right)}^{ac}\right)={dk}_{T_0}.F\left({e\left(g_1,g_2\right)}^{yc}\right)={dk}_{T_0}.F\left(Y^{s_1}\right)$, which implies that $s_1=c$.
Furthermore, $\mathcal{S}_0$ selects $s_1^\prime,s_1^{\prime\prime},s_2,s_2^{\prime\prime}\in_RZ_p^\ast$, and computes ${\widetilde{C}}_{w_\nu^\ast}^\prime=1.Y^{s_2}$, $C_{1,w_\nu^\ast}^\prime=g_2^{s_2^{\prime\prime}}$, ${\widehat{C}}_{1,w_\nu^\ast}^\prime=g_1^{s_2-s_2^{\prime\prime}}$, $C_{\Delta,w_\nu^\ast}=Y^{s_1^\prime}$, ${\widehat{C}}_{0,w_\nu^\ast}=g_1^{s_1^\prime}$, $C_{1,w_\nu^\ast}=g_2^{s_1^{\prime\prime}}$, ${\widehat{C}}_{1,w_\nu^\ast}=g_1^{s_1-s_1^{\prime\prime}}=\left(g^c.g^{-s^{\prime\prime}}\right)^\omega$.
$\mathcal{S}_0$ also picks $\{\sigma_{i,\Delta},\sigma_{i,0},\sigma_{i,1},\sigma_{i,0}^\prime,\sigma_{i,1}^\prime\in_RG|1\le i\le n\}$ such that $\prod_{i=1}^{n}\sigma_{i,\Delta}=\prod_{i=1}^{n}\sigma_{i,0}=\prod_{i=1}^{n}\sigma_{i,1}=\prod_{i=1}^{n}\sigma_{i,0}^\prime=\prod_{i=1}^{n}\sigma_{i,1}^\prime=1_\mathbb{G}$, and computes $[C_{i,t,\Delta,w_\nu^\ast},C_{i,t,0,w_\nu^\ast},{\widehat{C}}_{i,t,0,w_\nu^\ast},C_{i,t,0,w_\nu^\ast}^\prime,{\widehat{C}}_{i,t,0,w_\nu^\ast}^\prime]$ as follows:

\begin{enumerate}
      \item If $v_{i,t}\in W_{\nu,i}^\ast$, then
        \begin{flalign*}
            &\left[
            \begin{matrix}
            \begin{split}
                C_{i,t,\mathrm{\Delta},w_\nu^\ast}&=\sigma_{i,\mathrm{\Delta}}.H(i||v_{i,t})^{s_1^\prime}=\sigma_{i,\mathrm{\Delta}}.g^{\tau_{i,t}.s_1^\prime},\\
                C_{i,t,0,w_\nu^\ast}&=\sigma_{i,0}.H(0||i||v_{i,t})^{s_1^{\prime\prime}}=\sigma_{i,0}.g_2^{a_{i,t}.s_1^{\prime\prime}},\\
                {\widehat{C}}_{i,t,0,w_\nu^\ast}&=\sigma_{i,1}.H(1||i||v_{i,t})^{({s_{1}-s}_{1}^{\prime\prime})}\\&=\sigma_{i,1}.g^{b_{i,t}.({s_{1}-s}_{1}^{\prime\prime})}\\&=\sigma_{i,1}.\left(g^c\right)^{b_{i,t}}.g^{-b_{i,t}.s_1^{\prime\prime}},\\
                C_{i,t,0,w_\nu^\ast}^\prime&=\sigma_{i,0}^\prime.H(0||i||v_{i,t})^{s_2^{\prime\prime}}=\sigma_{i,0}^\prime.g_2^{a_{i,t}.s_2^{\prime\prime}},\\
                {\widehat{C}}_{i,t,0,w_\nu^\ast}^\prime&=\sigma_{i,1}^\prime.H(1||i||v_{i,t})^{{(s_2-s}_2^{\prime\prime})}\\&=\sigma_{i,1}^\prime.g^{b_{i,t}.{(s_2-s}_2^{\prime\prime})}
            \end{split}
            \end{matrix}
            \right].&
        \end{flalign*}

      \item If $v_{i,t}\notin W_{\nu,i}^\ast$, then $[C_{i,t,\mathrm{\Delta},w_\nu^\ast},C_{i,t,0,w_\nu^\ast},{\widehat{C}}_{i,t,0,w_\nu^\ast},C_{i,t,0,w_\nu^\ast}^\prime,{\widehat{C}}_{i,t,0,w_\nu^\ast}^\prime]$ are chosen randomly from $\mathbb{G}$.\\

\end{enumerate}

\noindent Finally, the simulator $\mathcal{S}_0$ returns the following challenge ciphertext of $M_\nu$ with respect to $A_\nu^\ast$:
\begin{flalign*}
    &{CT}_{cloud}^{A_\nu^\ast}=&\\
    &\left(
    \begin{matrix}
    \begin{split}
    	&U_0,U_1,V,
        {\widetilde{C}}_{w_\nu^\ast}^\prime,C_{1,w_\nu^\ast}^\prime,{\widehat{C}}_{1,w_\nu^\ast}^\prime,\\
        &\{\{
            \begin{matrix}
            \begin{split}
                &C_{i,t,0,w_\nu^\ast}^{\prime},{\widehat{C}}_{i,t,0,w_\nu^\ast}^{\prime}
            \end{split}
            \end{matrix}
        \}_{1\le t\le n_i}\}_{1\le i\le n},\\
        &{\widetilde{C}}_{w_\nu^\ast},C_{\Delta,w_\nu^\ast},{\widehat{C}}_{0,w_\nu^\ast},C_{1,w_\nu^\ast},{\widehat{C}}_{1,w_\nu^\ast},\\
        &\{\{
            \begin{matrix}
            \begin{split}
                &C_{i,t,\Delta,w_\nu^\ast},C_{i,t,0,w_\nu^\ast},{\widehat{C}}_{i,t,0,w_\nu^\ast}
            \end{split}
            \end{matrix}
        \}_{1\le t\le n_i}\}_{1\le i\le n}
    \end{split}
    \end{matrix}
    \right).&
\end{flalign*}

\textbf{Phase 2:} The adversary $\mathcal{A}$ continues to query the oracles as in \textbf{Phase 1}.

\textbf{Guess:} The adversary $\mathcal{A}$ guesses a bit $\nu^\prime$ as the value of $\nu$. $\mathcal{A}$ wins the game if $\nu^\prime=\nu$. When $\mathcal{A}$ wins the game, $\mathcal{S}_0$ returns $1$ and otherwise returns $0$ to the DBDH challenger. 
Indicate that if ${Z=e\left(g,g\right)}^{abc}$, then $\mathcal{A}$ is in game $G_0^\prime$, and otherwise, $Z$ is a random element in $\mathbb{G}_T$ and $\mathcal{A}$ is in game $G_0$. Therefore the advantage of $\mathcal{S}_0$ in the DBDH game is equal to $\epsilon_0$, where $\epsilon_0$ is the difference between the advantages of $\mathcal{A}$ to win the game $G_0^\prime$ and the game $G_0$. Finally, with respect to the DBDH assumption we have $\left|Pr\left[\mathcal{E}_0^\prime\right]-Pr\left[\mathcal{E}_0\right]\right|=\epsilon_0\le\epsilon_{DBDH}$.\\

\textbf{Lemma 3.} Under the D-Linear assumption, the difference between advantages of $\mathcal{A}$ in games $G_{h-1}$ and $G_h$ is negligible for $1\le h\le h_{max}$, such that
$|Pr\left[\mathcal{E}_{h-1}\right]-Pr\left[\mathcal{E}_h\right]|\le\epsilon_{DL}$.

\textbf{Proof.} We show that if $\epsilon_h=\left|Pr\left[\mathcal{E}_{h-1}\right]-Pr\left[\mathcal{E}_h\right]\right|$ is not negligible, then there exists a simulator $\mathcal{S}_h$ that can break the D-Linear assumption. To construct $\mathcal{S}_h$, suppose that it is given a D-Linear instance $[g,g^{z_1},g^{z_2},Z,g^{z_2z_4},g^{z_3+z_4}]$ by the challenger where $Z\in\mathbb{G}$ (note that this D-Linear assumption is equivalent to that of Section \ref{section:complexity_assumptions} \cite{11}). At the other side, $\mathcal{S}_h$ plays the role of a challenger for the adversary $\mathcal{A}$. In this manner, $\mathcal{S}_h$ will be able to win the D-Linear game with the non-negligible advantage $\epsilon_h$ by exploiting $\mathcal{A}$. 
Accordingly, the simulator $\mathcal{S}_h$ acts as follows:

\textbf{Init:} The adversary $\mathcal{A}$ gives $\mathcal{S}_h$ two challenge access policies $A_0^\ast=W_0^\ast=\left[W_{0,1}^\ast,\ldots,W_{0,n}^\ast\right]$ and $A_1^\ast=W_1^\ast=\left[W_{1,1}^\ast,\ldots,W_{1,n}^\ast\right]$, and a timestamp $T_l$. Afterward, $\mathcal{S}_h$ selects a random bit $\nu\in\{0,1\}$.
Suppose that the ciphertext components $\{C_{i_h,t_h,\mathrm{\Delta},w_\nu^\ast},C_{i_h,t_h,0,w_\nu^\ast},{\widehat{C}}_{i_h,t_h,0,w_\nu^\ast},C_{i_h,t_h,0,w_\nu^\ast}^\prime,{\widehat{C}}_{i_h,t_h,0,w_\nu^\ast}^\prime\}$ that are generated in a routine manner in game $G_{h-1}$, are chosen randomly from $\mathbb{G}$ in game $G_h$.
Also, without loss of generality, assume that $(v_{i_h,t_h}\notin W_{0,i_h}^\ast\land v_{i_h,t_h}\in W_{1,i_h}^\ast)$.
Thus, we proceed assuming $\nu=1$.

\textbf{Setup:} The simulator $\mathcal{S}_h$ selects $RK,S_{T_0},\omega,{mk}_0,{mk}_1,sk\in_RZ_p$, and $g_3,g_4\in_R\mathbb{G}$, sets $g_1=g^{z_1}$ and $g_2=g^{z_2}$, and computes $Q_0=g_3^{sk}$, ${PP}_0={e\left(g_3,g_4\right)}^{{mk}_0}$, ${PP}_1=g_3^{{mk}_1}$. 
Afterwards, $\mathcal{S}_h$ selects $y\in_RZ_p$ and computes $Y={e(g_1,g_2)}^y$. 
Finally, $\mathcal{S}_h$ sends the system public key $PK=\langle g,g_1,g_2,g_3,g_4,Y \rangle$ and the data public parameter $PP=(Q_0,{PP}_0,{PP}_1)$ to $\mathcal{A}$.

\textbf{Phase 1:} The simulator $\mathcal{S}_h$ answers  $\mathcal{A}$'s queries by simulating $\mathcal{O}_{Hash}$ and $\mathcal{O}_{AttrKeyGen}$ as follows:

\begin{itemize}
        \item[$\bullet$] \textbf{Hash Query $\mathcal{O}_{Hash}\left(\mathcal{M}\right)$:} Firstly, the simulator $\mathcal{S}_h$ selects an empty list $\mathcal{L}_H$. When the adversary $\mathcal{A}$ queries the random oracle $\mathcal{O}_{Hash}\left(.\right)$ for an input $\mathcal{M}$, $\mathcal{S}_h$ checks in the list $\mathcal{L}_H$ if the value of $H(\mathcal{M})$ has been defined. If there exists a record in $\mathcal{L}_H$ associated to the queried point $\mathcal{M}$, $\mathcal{S}_h$ returns the previously defined value. 
Otherwise, $\mathcal{S}_h$ selects $\{\{\tau_{i,t},a_{i,t},b_{i,t}\in_RZ_p\}_{1\le t\le n_i}\}_{1\le i\le n}$, and calculates the output as follows:
            \begin{enumerate}[label=\arabic*)]
              \item For $\mathcal{M}=(i||v_{i,t}$), $\mathcal{S}_h$ sets $\alpha=\tau_{i,t}$, returns $H(\mathcal{M})=g^\alpha$ if $v_{i,t}\in W_{\nu,i}^\ast$, or $H(\mathcal{M})=\left(g^{z_1}\right)^\alpha$ if $v_{i,t}\notin W_{\nu,i}^\ast$.
              \item For $\mathcal{M}=(0||i||v_{i,t})$, $\mathcal{S}_h$ sets $\alpha=a_{i,t}$, returns $H(\mathcal{M})=\left(g^{z_2}\right)^\alpha$ if $v_{i,t}\in W_{\nu,i}^\ast$, or $H(\mathcal{M})=g^\alpha$ if $v_{i,t}\notin W_{\nu,i}^\ast$.
              \item For $\mathcal{M}=(1||i||v_{i,t})$, $\mathcal{S}_h$ sets $\alpha=b_{i,t}$, returns $H(\mathcal{M})=\left(g^{z_1}\right)^\alpha$ if $v_{i,t}\in W_{\nu,i}^\ast$, or $H(\mathcal{M})=g^\alpha$ if $v_{i,t}\notin W_{\nu,i}^\ast$.
            \end{enumerate}

        \noindent Finally, $\mathcal{S}_h$ adds $\langle \mathcal{M},\alpha,H(\mathcal{M}) \rangle$ to the list $\mathcal{L}_H$.

        \item[$\bullet$] \textbf{AttrKeyGen Query $\mathcal{O}_{AttrKeyGen}\left(L\right)$:} When the adversary $\mathcal{A}$ queries the random oracle $\mathcal{O}_{AttrKeyGen}\left(.\right)$ for an attribute list $L=[L_1,L_2,\ldots,L_n]$ where $L_i=v_{i,k_i}$ and with the restriction that $(L\nvDash A_0^\ast\land L\nvDash A_1^\ast)$, the simulator $\mathcal{S}_h$ firstly selects $r_1,r_2,\ldots,r_{n}\in_RZ_p$ such that $\sum_{i=1}^{n}r_i=y$. 
Besides, $\mathcal{S}_h$ selects $r,\lambda,\hat{\lambda}\in_RZ_p$ and $\left\{{\hat{r}}_i\in_RZ_p\right\}_{1\le i\le n}$, sets $\hat{r}=\sum_{i=1}^{n}{\hat{r}}_i$, and computes $D_{\mathrm{\Delta},0}=g_1^r$,${\widehat{D}}_{\mathrm{\Delta},0}=g_2^{y-\hat{r}}$,$D_0=g_2^\lambda$,${\widehat{D}}_0=g_1^{\hat{\lambda}}$. Afterwards, assuming $L_i=v_{i,k_i}$ for $1\le i\le n$, $\mathcal{S}_h$ computes the rest of the attribute secret key components as follows:

            \begin{enumerate}
            \item If $v_{i,k_i}\in W_{\nu,i}^\ast$, then
                    \begin{flalign*}
                        &\left[
                        \begin{matrix}
                        \begin{split}
                            &D_{\mathrm{\Delta},i}=g_2^{{\hat{r}}_i}.g^{r.\tau_{i,k_i}},
                            D_{i,1}=g_1^{r_i}.\left(g^{z_2}\right)^{\lambda a_{i,k_i}},\\
                            &{\widehat{D}}_{i,1}=g_2^{r_i}.\left(g^{z_1}\right)^{\hat{\lambda}b_{i,k_i}}
                        \end{split}
                        \end{matrix}
                        \right].&
                    \end{flalign*}
         \item If $v_{i,k_i}\notin W_{\nu,i}^\ast$, then
                      \begin{flalign*}
                        &\left[
                        \begin{matrix}
                        \begin{split}
                            &D_{\mathrm{\Delta},i}=g_2^{{\hat{r}}_i}.\left(g^{z_1}\right)^{r.\tau_{i,k_i}},
                            D_{i,1}=g_1^{r_i}.g^{\lambda a_{i,k_i}},\\
                            &{\widehat{D}}_{i,1}=g_2^{r_i}.g^{\hat{\lambda}b_{i,k_i}}
                        \end{split}
                        \end{matrix}
                        \right].&
                    \end{flalign*}
            \end{enumerate}

\noindent Finally, the following attribute secret key is returned:
${SK}_L=\langle D_0, {\widehat{D}}_0, D_{\mathrm{\Delta},0}, {\widehat{D}}_{\mathrm{\Delta},0}, {\{ D_{\mathrm{\Delta},i}, D_{i,1}, {\widehat{D}}_{i,1} \}}_{1\le i\le n} \rangle$.

        \item[$\bullet$] \textbf{Reencrypt Query $\mathcal{O}_{Reencrypt}\left(T_i,{CT}_{cloud}^A\right)$:} 	The simulator $\mathcal{S}_h$ proceeds as in \textbf{Lemma 2}.

\end{itemize}

\textbf{Challenge:} The adversary $\mathcal{A}$ sends two different equal-length messages $M_0$ and $M_1$ to the simulator $\mathcal{S}_h$. 
This latter sets $V$ and ${\widetilde{C}}_{w_\nu^\ast}$ to be random elements in $\mathbb{G}_T$ and $\mathbb{G}$, respectively. 
Afterwards, $\mathcal{S}_h$ selects $r_d\in_RZ_p^\ast$, and sets $U_0=g_3^{r_d}$, $U_1=g_3^{r_d.sk}$, ${dk}_{T_0}=g_4^{{mk}_0}.g_3^{sk.{mk}_1}.g_3^{{mk}_1.S_{T_0}}$. 
Besides, $\mathcal{S}_h$ selects $s_1^\prime,s_2,s_2^{\prime\prime}\in_RZ_p$, and computes ${\widetilde{C}}_{w_\nu^\ast}^\prime=1.Y^{s_2}$, $C_{1,w_\nu^\ast}^\prime=g_2^{s_2^{\prime\prime}}$, ${\widehat{C}}_{1,w_\nu^\ast}^\prime=g_1^{s_2-s_2^{\prime\prime}}$, $C_{\Delta,w_\nu^\ast}=Y^{s_1^\prime}$, ${\widehat{C}}_{0,w_\nu^\ast}=g_1^{s_1^\prime}$. 
Moreover, $\mathcal{S}_h$ sets $C_{1,w_\nu^\ast}=g^{z_2z_4}=g_2^{s_1^{\prime\prime}}$ and ${\widehat{C}}_{1,w_\nu^\ast}=Z=g^{z_1z_3}=g_1^{s_1-s_1^{\prime\prime}}$ which respectively imply that $s_1^{\prime\prime}=z_4$ and $s_1=z_3+z_4$. 
Then, $\mathcal{S}_h$ selects $\{\sigma_{i,\Delta},\sigma_{i,0},\sigma_{i,1},\sigma_{i,0}^\prime,\sigma_{i,1}^\prime\in_R\mathbb{G}|1\le i\le n\}$ such that $\prod_{i=1}^{n}\sigma_{i,\Delta}=\prod_{i=1}^{n}\sigma_{i,0}=\prod_{i=1}^{n}\sigma_{i,1}=\prod_{i=1}^{n}\sigma_{i,0}^\prime=\prod_{i=1}^{n}\sigma_{i,1}^\prime=1_\mathbb{G}$, and generates the ciphertext components $\{\{C_{i,t,\Delta,w_\nu^\ast},\ C_{i,t,0,w_\nu^\ast},{\widehat{C}}_{i,t,0,w_\nu^\ast},C_{i,t,0,w_\nu^\ast}^\prime,{\widehat{C}}_{i,t,0,w_\nu^\ast}^\prime\}_{1\le t\le n_i}\}_{1\le i\le n}$ as in game $G_{h-1}$ while the components $\{C_{i_h,t_h,\Delta,w_\nu^\ast},C_{i_h,t_h,0,w_\nu^\ast},{\widehat{C}}_{i_h,t_h,0,w_\nu^\ast},C_{i_h,t_h,0,w_\nu^\ast}^\prime,{\widehat{C}}_{i_h,t_h,0,w_\nu^\ast}^\prime\}$ are computed as follows:
\begin{flalign*}
    &\left[
    \begin{matrix}
    \begin{split}
        C_{i_h,t_h,\mathrm{\Delta},w_\nu^\ast}&=\sigma_{i_h,\mathrm{\Delta}}.H(i_h||v_{i_h,t_h})^{s_1^\prime}=\sigma_{i_h,\mathrm{\Delta}}.g^{\tau_{i_h,t_h}.s_1^\prime},\\
        C_{i_h,t_h,0,w_\nu^\ast}&=\sigma_{i_h,0}.H(0||i_h||v_{i_h,t_h})^{s_1^{\prime\prime}}=\sigma_{i_h,0}.g_2^{a_{i_h,t_h}.s_1^{\prime\prime}}\\&=\sigma_{i_h,0}.\left(g^{z_2z_4}\right)^{a_{i_h,t_h}},\\
        {\widehat{C}}_{i_h,t_h,0,w_\nu^\ast}&=\sigma_{i_h,1}.H(1||i_h||v_{i_h,t_h})^{({s_1-s}_1^{\prime\prime})}\\&=\sigma_{i_h,1}.{(g_1)}^{b_{i_h,t_h}.({s_1-s}_1^{\prime\prime})}\\&=\sigma_{i_h,1}.(g^{z_1z_3})^{b_{i_h,t_h}}=\sigma_{i_h,1}.Z^{b_{i_h,t_h}},\\
        C_{i_h,t_h,0,w_\nu^\ast}^\prime&=\sigma_{i_h,0}^\prime.H(0||i_h||v_{i_h,t_h})^{s_2^{\prime\prime}}=\sigma_{i_h,0}^\prime.g_2^{a_{i_h,t_h}.s_2^{\prime\prime}},\\
        {\widehat{C}}_{i_h,t_h,0,w_\nu^\ast}^\prime&=\sigma_{i_h,1}^\prime.H(1||i_h||v_{i_h,t_h})^{{(s_2-s}_2^{\prime\prime})}\\&=\sigma_{i_h,1}^\prime.g_1^{b_{i_h,t_h}.{(s_2-s}_2^{\prime\prime})}\\
    \end{split}
    \end{matrix}
    \right].&
\end{flalign*}


\textbf{Phase 2:} The adversary $\mathcal{A}$ continues to query the oracles as in \textbf{Phase 1}.

\textbf{Guess:} The adversary $\mathcal{A}$ guesses a bit $\nu^\prime$ as the value of $\nu$. $\mathcal{A}$ wins the game if $\nu^\prime=\nu$. When $\mathcal{A}$ wins the game, $\mathcal{S}_h$ returns $1$ and otherwise returns $0$ to the D-Linear challenger.
Indicate that if $Z=g^{z_1z_3}$, then $\mathcal{A}$ is in game $G_{h-1}$, and otherwise, $Z$ is a random element in $\mathbb{G}$ and $\mathcal{A}$ is in game $G_h$.
Therefore the advantage of $\mathcal{S}_h$ in the D-Linear game is equal to $\epsilon_h$, where $\epsilon_h$ is the difference between the advantages of $\mathcal{A}$ to win the game $G_{h-1}$ and the game $G_h$. Finally, with respect to the D-Linear assumption we have $\left|Pr\left[\mathcal{E}_{h-1}\right]-Pr\left[\mathcal{E}_h\right]\right|=\epsilon_h\le\epsilon_{DL}$ for $1\le h\le h_{max}$.

\begin{table*}[t]
  \centering 
  \vspace{-5mm}
    \caption{STORAGE AND EFFICIENCY COMPARISON.}
    \label{tab:table_performance_compare}
    \setlength\extrarowheight{-5pt}
    \resizebox{\textwidth}{!}{
        \begin{tabular}{l l l l l l}
          \toprule 
          \multicolumn{1}{l}{\textbf{Scheme}}&
          \multicolumn{1}{l}{
            \begin{tabular}{c}
                \textbf{System public}\\
                \textbf{key size}\\
            \end{tabular}
          } &
          \multicolumn{1}{l}{
            \begin{tabular}{c}
              \textbf{Attribute}\\
              \textbf{secret key size}\\
            \end{tabular}
          } &
          \multicolumn{1}{l}{
            \begin{tabular}{c}
              \textbf{Ciphertext}\\
              \textbf{size}\\
            \end{tabular}
          } &
          \multicolumn{1}{l}{
            \begin{tabular}{c}
              \textbf{Encryption}\\
              \textbf{cost}\\
            \end{tabular}
          } &
          \multicolumn{1}{l}{
            \begin{tabular}{c}
              \textbf{Decryption cost}\\
              \textbf{(decryption phase)}\\
            \end{tabular}
          }\\
          \midrule 
            \cite{5}  & $3|\mathbb{G}|+1|\mathbb{G}_T|$ & $(3n+4)|\mathbb{G}|$ & $(3N+3)|\mathbb{G}|+2|\mathbb{G}_T|$ & $M_\mathbb{G}(3n+1)+2E_{\mathbb{G}_T}+E_\mathbb{G}(3n+3)+R_\mathbb{G}(3N)$ & $M_\mathbb{G}(2n)+4M_{\mathbb{G}_T}+4P$\\
          \midrule 
            \cite{31} & $(3N+4)|\mathbb{G}|+1|\mathbb{G}_T|$ & $(4n+4)|\mathbb{G}|$ & $(3N+3)|\mathbb{G}|+2|\mathbb{G}_T|$ & $M_\mathbb{G}(n+1)+2E_{\mathbb{G}_T}+E_\mathbb{G}(3n+3)+R_\mathbb{G}(3N)$ & $M_{\mathbb{G}_T}(3n+1)+P(3n+1)$\\
          \midrule 
            \cite{32} & $(2a+11)|\mathbb{G}|$ &	$(5n+2)|\mathbb{G}|$ & $(6l+3)|\mathbb{G}|$ & $M_\mathbb{G}(2l+2)+1E_{\mathbb{G}_T}+E_\mathbb{G}(8l+4)+1P$ 
            &	
            \begin{tabular}{l} 
            	$M_{\mathbb{G}_T}(5l+3)$\\ 
            	$+E_{\mathbb{G}_T}(l+2)+P(6l+3)$
            \end{tabular}\\
          \midrule 
            \cite{48} & $(n+5)|\mathbb{G}|+1|\mathbb{G}_T|$ & $3|\mathbb{G}|+(n+2)|\mathbb{Z}_p|$ & $(n+4)|\mathbb{G}|+1|\mathbb{G}_T|+(n+2)|\mathbb{Z}_p|$ 
            & 
            \begin{tabular}{l l}
                   DO (offline): & $M_\mathbb{G}(n)+1E_{\mathbb{G}_T}+E_\mathbb{G}(2n+2)$\\
                   DO (online): & $1M_{\mathbb{G}_T}$\\
            \end{tabular}
            & 
            \begin{tabular}{l}
            	$M_{\mathbb{G}}(n)+3M_{\mathbb{G}_T}$\\ $+E_\mathbb{G}(n+3)+3P$ \\
            \end{tabular}\\
          \midrule 
            HUAP & $5|\mathbb{G}|+1|\mathbb{G}_T|$
            	 & $(3n+4)|\mathbb{G}|$
            	 &
            	 \begin{tabular}{l l}
            	 	\begin{tabular}{l} message\\ ciphertext \end{tabular}
            	 	: & 	$2|\mathbb{G}|+1|\mathbb{G}_T|$\\
            	 	\begin{tabular}{l} policy\\ ciphertext \end{tabular}
            	 	: & \begin{tabular}{l}
            	 			$(5mN+6m)|\mathbb{G}|$\\
            	 			$+(2m)|\mathbb{G}_T|$
            	 		\end{tabular}\\
	             \end{tabular}
            	 &
            \begin{tabular}{l l}
                   Dev (offline): & $1E_{\mathbb{G}_T}+2E_\mathbb{G}$\\
                   Dev (online): & $1M_{\mathbb{G}_T}$\\
                   DO: & 	\begin{tabular}{l}
                   			$M_\mathbb{G}(5mn+m)+E_{\mathbb{G}_T}(3m)$\\
                   			$+E_\mathbb{G}(5mn+5m)+R_\mathbb{G}(5mN)$
                   			\end{tabular}\\
            \end{tabular}
            & 
            \begin{tabular}{l}
	            $M_\mathbb{G}(4n+2)$\\ 
	            $+9M_{\mathbb{G}_T}+10P$
            \end{tabular}
            \\	
          \bottomrule 
        \end{tabular}
    }
\end{table*}

\section{PERFORMANCE EVALUATION}
In this section, we evaluate the effectiveness of our proposed scheme by comparing its computational and storage complexities with schemes presented in \cite{5,31,32,48}. The main reason for considering the schemes for comparison  is  that they are similar to our proposed scheme in several aspects such as functional capabilities or cryptographic algorithms.
Table \ref{tab:table_performance_compare} summarizes the comparison results.
In this table, $n, m, l,$ and $a$ denote the total number of attributes in the universe, the number of AND-gates in an access policy, the number of rows in an LSSS access policy matrix, and the maximum number of users in the system, respectively. $N=\sum_{i=1}^{n}n_i$ indicates the total number of possible values of all attributes. 
Also, $M, E, P$, and $R$ represent a modular multiplication, a modular exponentiation, a bilinear pairing, and a random element selection, respectively.

\begin{figure}
    \centering

\comment{
	\subcaptionbox{Encryption.}[.45\columnwidth][c]
	{%
		\includegraphics[width=.45\columnwidth]{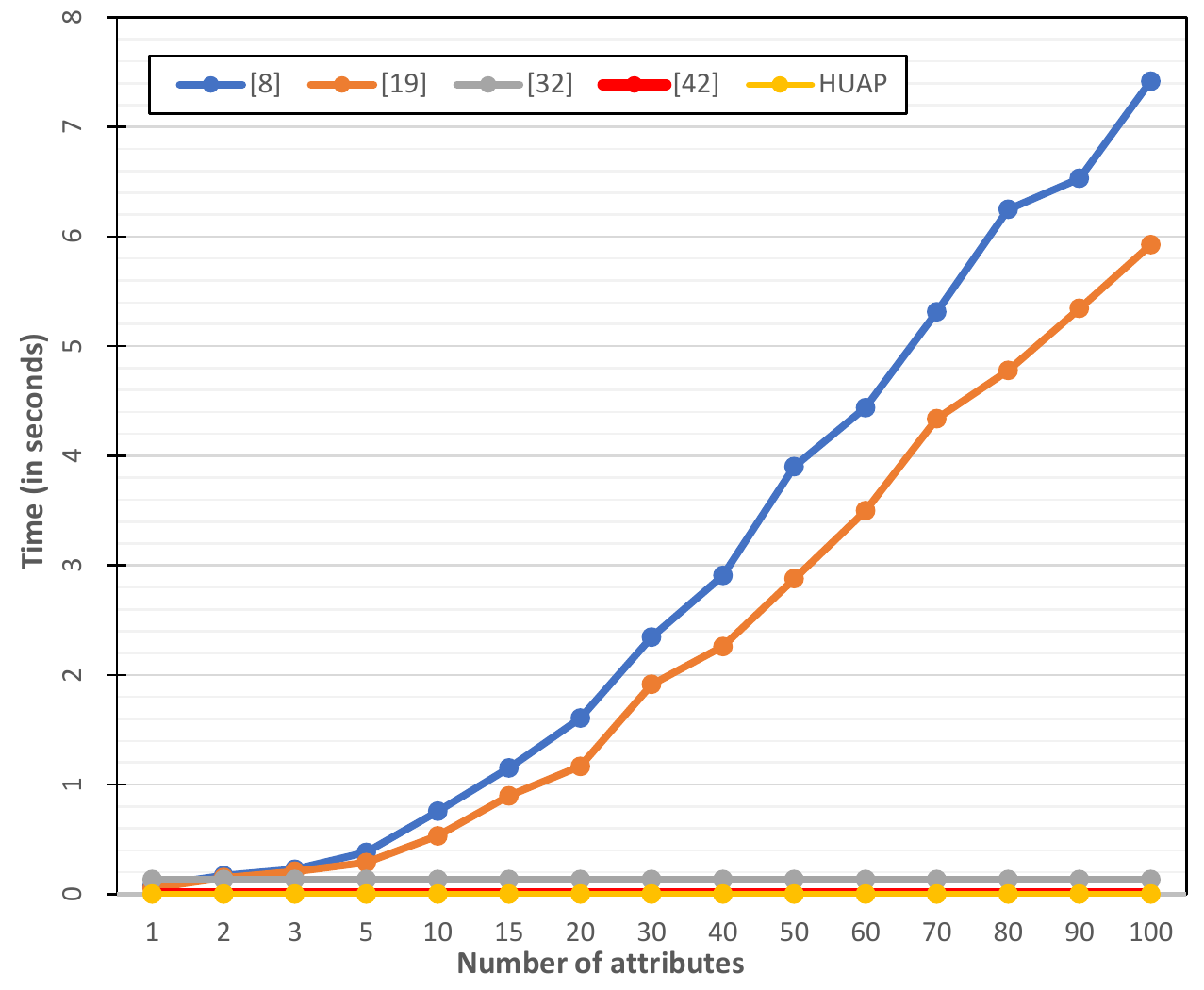}
	}\quad
	\subcaptionbox{Decryption.}[.45\columnwidth][c]{%
		\includegraphics[width=.45\columnwidth]{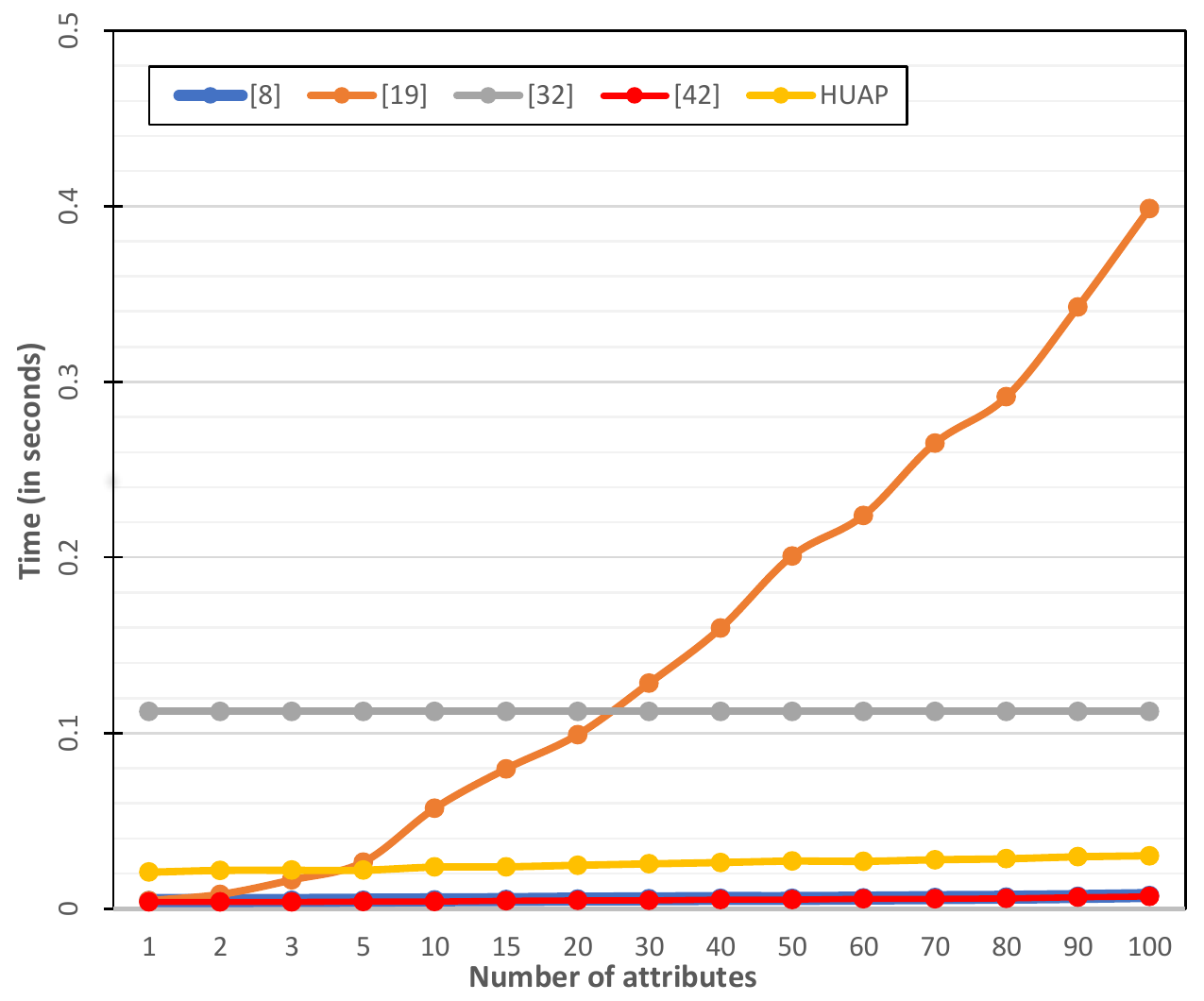}}\quad
}

\comment{
	\subcaptionbox{Offline encryption time}[.45\columnwidth][c]{%
		\includegraphics[width=.45\columnwidth]{Figure_Encryption.pdf}}\quad
	\subcaptionbox{Online encryption time}[.45\columnwidth][c]{%
		\includegraphics[width=.45\columnwidth]{Figure_Decryption.pdf}}\quad
}

\comment{
    \begin{subfigure}[b]{0.49\columnwidth}
     \centering
     \includegraphics[width=\textwidth]{Figure_Encryption.pdf}
     \caption{Encryption.}
     \label{fig:Enc_compare}
    \end{subfigure}
    \begin{subfigure}[b]{0.49\columnwidth}
	 \centering
     \includegraphics[width=\textwidth]{Figure_Decryption.pdf}
     \caption{Decryption.}
     \label{fig:Dec_compare}
    \end{subfigure}
}

    \begin{subfigure}[b]{0.9\columnwidth}
     \centering
     \includegraphics[width=0.9\columnwidth]{Figure_Encryption.pdf}
     \caption{Encryption.}
     \label{fig:Enc_compare}
    \end{subfigure}
    \begin{subfigure}[b]{0.9\columnwidth}
	 \centering
     \includegraphics[width=0.9\columnwidth]{Figure_Decryption.pdf}
     \caption{Decryption.}
     \label{fig:Dec_compare}
    \end{subfigure}

    \caption{Performance Comparison.}
    \label{fig:Encryption_Decryption_compare}
\end{figure}

We have implemented the simulation experiment on a virtual machine equipped with Intel Core i7-3632QM CPU (2 core 2.20 GHz) and 2 GB memory running Linux Kernel 5.4.0. The experiment is implemented with PBC library of version 0.5.14 for underlying cryptographic operations. 
The evaluation results of executing encryption and decryption algorithms are presented in Figure \ref{fig:Encryption_Decryption_compare}, where we have set $l=10$, 
$a=10$, 
and $n_i=10$ for $1\le i\le n$.
To be specific, we have compared the cost of online encryption that is performed by Dev in our scheme, with the cost of encryption in other schemes.
Moreover, as the number of random element selections in \cite{5}, \cite{31}, and HUAP appreciably affects the encryption cost, it is enumerated in Table \ref{tab:table_performance_compare} for these schemes.
From Figure \ref{fig:Enc_compare}, the online encryption time in our scheme is constant as the number of attributes in the universe is increased.
In the meanwhile, from Figure \ref{fig:Dec_compare}, in our scheme, growing the total number of attributes in the universe does not change the decryption time significantly, as the number of pairing operations in the decryption algorithm is constant.
Note that although the complexity of the encryption/decryption algorithm in\cite{32} remains at a constant value when the total number of attributes in the universe is increased, it grows linearly with respect to the number of rows in the access policy matrix.


\section{CONCLUSION}

In this paper, a ciphertext-policy attribute-based access control scheme has been proposed.
In the proposed scheme the access policies are hidden and hence unauthorized data users cannot learn which attribute set  satisfies an access policy.
The scheme also enables data owners to efficiently outsource a major part of the access policy update process to a cloud service provider. In particular, the process does not require generating a new re-encryption key.
Moreover, to reduce the computational cost for resource-constrained devices, this scheme divides the encryption algorithm into two offline and online phases.
Furthermore, in the proposed scheme, the decryption process is very fast and requires only a constant number of bilinear pairing operations. 
The proposed scheme is proven to be secure in the random oracle model. Our simulation results indicate that our proposed scheme effectively decreases computational overhead at the data collector device and data user sides.

\section*{Acknowledgement}
This work was partially supported by Iran National Science Foundation (INSF) under Grant No. 96.53979.







\bibliographystyle{elsarticle-num-names}
\bibliography{HUAP_References} 

\comment{
\begin{wrapfigure}{l}{25mm} 
    \includegraphics[width=1in,height=1.25in,clip,keepaspectratio]{example-image}
  \end{wrapfigure}\par
  \textbf{Author A} is a well-known author in the field of the journal scope. His/Her research interests include interest 1, interest 2.\par
}

\comment{
\InsertBoxL{0}{\includegraphics[width=25mm,height=25mm,keepaspectratio]{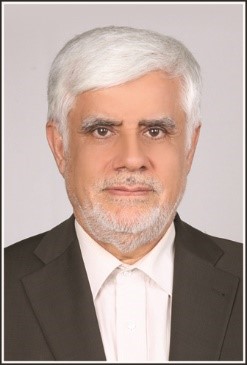}}[-1]
\textbf{Author Alpha} \Blindtext[1][13]
}

\comment{
\authorbibliography[scale=0.3,wraplines=10,overhang=40pt,imagewidth=4cm,imagepos=r]{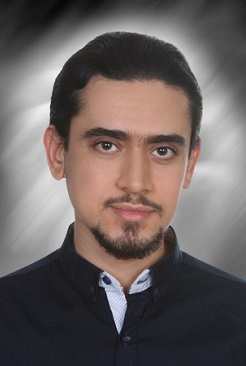}{Mostafa Chegenizadeh}{was born 16 May 1995 in Dezful, Iran. He received the B.Sc. and M.Sc. degrees in Electrical Engineering from Sharif University of Technology, Tehran, Iran, in 2017 and 2019, respectively. His current research interests include cloud security, Internet of Things security, and cryptographic protocols.}
}

\comment{
\bio{}
Author biography. Author biography. Author biography.
\endbio
}

\subsection*{  } 

\begin{wrapfigure}{l}{1.0in} 
	\vspace{-15pt}
    \includegraphics[height=1.5in]{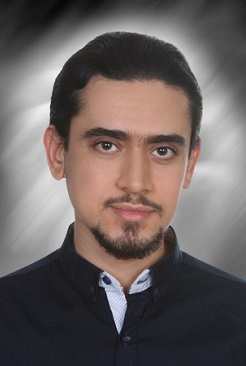}
\end{wrapfigure}
\noindent  \textbf{Mostafa Chegenizadeh} was born on May 16, 1995, in Dezful, Iran. He received the B.Sc. and M.Sc. degrees in Electrical Engineering from Sharif University of Technology, Tehran, Iran, in 2017 and 2019, respectively. His current research interests include cloud security, Internet of Things security, and cryptographic protocols.

\subsection*{  } 

\begin{wrapfigure}{l}{1.0in} 
	\vspace{-5pt}
    \includegraphics[height=1.5in]{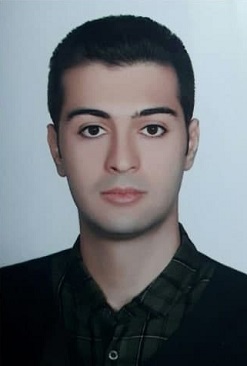}
\end{wrapfigure}
\noindent  \textbf{Mohammad Ali} received the B.Sc. degree in applied mathematics from Shahed University, Theran, Iran, in 2014, and the M.Sc. and Ph.D. degrees in applied mathematics from Amirkabir University of Technology, Tehran, Iran, in 2016 and 2020, respectively. His fields of interests are cryptography and cloud computing.

\subsection*{  } 

\begin{wrapfigure}{l}{1.0in} 
	\vspace{-15pt}
    \includegraphics[height=1.5in]{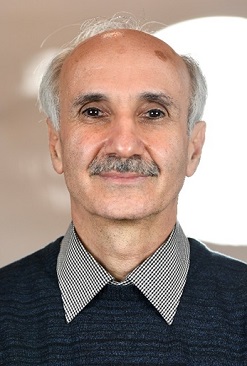}
\end{wrapfigure}
\noindent  \textbf{Javad Mohajeri} is an Assistant Professor with the Electronics Research Institute, Sharif University of Technology, Tehran, Iran, where he is an Adjunct Assistant Professor with the Electrical Engineering Department. He has authored or co-authored 3 books and 116 research articles in refereed journals/conferences. His current research interests include data security, and the design and analysis of cryptographic protocols and algorithms. Javad is a Founding Member of the Iranian Society of Cryptology. Also, he has been committee program chair of the second International ISC Conference on Information Security and Cryptology, and committee program member of the third – 17th of this conference.

\subsection*{  } 

\begin{wrapfigure}{l}{1.0in} 
	\vspace{-15pt}
    \includegraphics[height=1.5in]{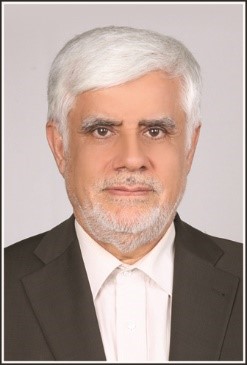}
\end{wrapfigure}
\noindent  \textbf{Mohammad Reza Aref} was born 19 December 1951 in Yazd, Iran. He received The B.Sc. degree from School of Electrical and Computer Engineering, University of Tehran, in 1975. The M.Sc. and Ph.D. degrees from Stanford University, Stanford, CA, USA, in 1976 and 1980, respectively. He Came back to Iran in 1980 and was actively engaged in academic and political affairs. He was a Faculty member of Isfahan University of Technology (1982-1997). He has been a Professor of Electrical Engineering at Sharif University of Technology Since 1997. He has published 439 technical papers in the field of Communication and Information Theory and Cryptography in international journals and conferences proceedings.

\end{document}